\documentclass[%
 reprint,
 superscriptaddress,
 amsmath,amssymb,
 aps,
 prb,
]{revtex4-1}

\usepackage[pdftex]{graphicx}
\usepackage[pdftex]{hyperref}
\usepackage{dcolumn}
\usepackage{bm}
\usepackage{color,soul}

\newlength{\mywidth}
\usepackage[normalem]{ulem}

\begin{document}

\title{The Vortex-Particle Magnus effect}

\author{Adam Griffin}
\email{a.griffin.1@warwick.ac.uk}
\affiliation{Mathematics Institute, The University of Warwick, Coventry, CV4 7AL, United Kingdom}
\author{Sergey Nazarenko}%
\affiliation{Institut de Physique de Nice, Universit\'e C\^ote d'Azur, CNRS, Nice, France}%
\author{Vishwanath Shukla}
\email{research.vishwanath@gmail.com}
\affiliation{Department of Physics, Indian Institute of Technology Kharagpur, Kharagpur - 721 302, India}%
\affiliation{Centre for Theoretical Studies, Indian Institute of Technology Kharagpur, Kharagpur - 721302, India}
\author{Marc-Etienne Brachet}
\affiliation{Laboratoire de Physique de l’Ecole normale supérieure, ENS, Université PSL, CNRS, Sorbonne}

\date{\today}

\begin{abstract}
Experimentalists use particles as tracers in liquid helium. The intrusive effects of particles on the dynamics of vortices remain poorly understood. We implement a study of how basic well understood vortex states, such as a propagating pair of oppositely signed vortices, change in the presence of particles by using a simple model based on the Magnus force. We focus on the 2D case, and compare the analytic and semi-analytic model with simulations of the Gross-Pitaevskii (GP) equation with particles modelled by dynamic external potentials. The results confirm that the Magnus force model is an effective way to approximate vortex-particle motion either with closed-form simplified solutions or with a more accurate numerically solvable ordinary differential equations (ODEs). Furthermore, we increase the complexity of the vortex states and show that the suggested semi-analytical model remains robust in capturing the dynamics observed in the GP simulations.

\end{abstract}

\maketitle


\section{Introduction}

Superfluidity occurs in a wide variety of systems both terrestrial, e.g., $^4$He-II, $^3$He-B, versatile Bose-Einstein condensates (BECs) of alkali atoms, exciton-polariton condensates \cite{Byrnes:2014aa} in laboratory experiments, and exotic astrophysical objects, such as neutron stars \cite{Baym:1992aa,Warszawski2012}. Moreover, a monochromatic light while passing through a non-linear media, e.g., photo-refractive crystals \cite{Michel:2018aa}, has been shown to exhibit a flow that is essentially superfluid. The unusual flow properties of these superfluid flows have held the attention of experimentalists and theorists alike. For example, helium II can sustain rotational motion only through formation of quantised vortices, wherein the circulation along paths enclosing vortices is restricted to multiples of $h/m_{4}$, where $h$ is the Planck's constant and $m_{4}$ is the mass of a $^4$He atom~\cite{Donnelly}. These quantised vortices in helium II are angstrom size in diameter, and they occur either as closed loops or filaments that must end at the boundary of the fluid.

Quantised vortices display rich dynamical behaviour~\cite{White14}: For example, in three dimensional (3D) rotating superfluid systems, above vortex nucleation threshold, the number of vortices increases with rotation speed, can form a vortex lattice. At still higher rotation  speed the system transitions to a turbulent state~\cite{TsubotaRotPRL}. It is a non-equilibrium state involving processes that span a broad range of length- and time-scales, and is characterised by the presence of a dynamic random tangle of interacting vortices~\cite{Feynman55,Nore97b,Nore97a}. These quantised vortices upon close approach can undergo reconnections~\cite{Bewley:2008aa,Schwarz85}, a topology changing process that further drives the system to a turbulent state. Reconnection events excite Kelvin waves on the vortices, and the non-linear interaction of these waves gives rise to a cascade process that transfers energy to smaller length scales, which is ultimately radiated as sound~\cite{Krstulovic12,Baggaley14,Fonda:2014aa,ShuklaTYGPRA19}. Turbulence in these superfluid systems can also be excited by stirring, shaking, moving objects, etc.~\cite{Maurer98,skrbek2009vibrating,Henn09}. Moreover, turbulent state can also be realised in two-dimensional (2D) superfluids\cite{Nazarenko06,Desbuquois:2012aa}, where point-like vortices move chaotically\cite{Neely2DQTPRL2013,Johnstone:2019aa}, and can organise to produce a large scale flows. The subject has seen a spurt of activity, both numerical and experimental, to find its universal features and provide a comparison with its classical counterpart~\cite{Bradley12,vmrnjp13,Shukla15}.

However, it must be emphasised the experimental study of the fundamental processes involved in superfluid turbulence is a difficult task and requires state of the art facilities~\cite{Donnelly,barenghiPNASRev2014experimental}. In particular, visualisation of the quantised vortices has been a challenge because of extremely low temperatures and small system sizes. The flow visualisation methods available for classical fluids are difficult to adapt to superfluid helium~\cite{guoPNASRev2014visualization}.

Use of particles to probe superfluid flows involving vortices was suggested in Ref.~\cite{Don_part}.  A great deal of information about quantised vortices, including their existence, in helium II has been obtained, in the past, by the use of moving ions~\cite{PhysRevLett.34.924}. More recently, solid-hydrogen particles were used to visualise quantised vortices in helium II at temperatures $\sim 2$\,K \cite{GBewley:2006aa,Duda2015}. These particles were also used to study vortex reconnection events~\cite{Bewley:2008aa} and Kelvin waves~\cite{Fonda:2014aa} on vortices in superfluid helium. Note that in these experiments, even though particles used were roughly $10^4$ times the vortex diameter, they managed to capture the essential physics.

Now much smaller particles in the form of metastable $^4 \rm{He}^*_{2}$ excimer molecules, $\sim100$\,nm in size, are available that can be used as vorticity tracers in the $T=0$ limit, in the absence of normal fluid; at temperatures above $1$\,K they act as tracers of the normal component \cite{Zmeev:2013aa}.

Notwithstanding the significant experimental progress in the use of particles to characterize superfluid turbulence, the exact level of intrusion of the particle on the vortex motion remains unclear. Therefore, it is important to explore and understand these issues by building simplified models to study particle-vortex dynamics both theoretically and numerically\cite{Berloff2001,Varga:2016aa}.

In this work, we focus on the Magnus force that acts on a particle trapped on a translating vortex. To do so, we make use of both the Gross-Pitaevskii equation (GPE) description of superfluids~\cite{Gross1961,Pit1961} and the suitably adapted classical treatment of point vortices in 2D. In our GPE description, we make use of a recently developed minimal model~\cite{shuklaPRA16,pandit2017pofoverview,Vish2017}, wherein we couple the equations of motion of particles with the GP classical wave function $\psi$. We demonstrate that the Magnus force description borrowed from the theory of ideal hydrodynamic flow works provides a good description of the dynamics of particle loaded vortices in superfluids. We first study a particle loaded vortex-antivortex pair configuration and carry out a systematic direct numerical simulations (DNS) of the GPE based minimal model. We compare these GPE DNS results with a Magnus force model that we have derived (see below), both in the simplified analytically tractable case of constant background flow and a more realistic situation wherein the background flow is allowed to vary in response the particle loaded vortices (or other external vortices). We then extend our study of the particle loaded vortex-antivortex pair to more complex vortex configurations, where: (i) each vortex is multiply charged; (ii) free external vortices are present in the neighbourhood. We will argue that under certain circumstances, the effects can be observed in superfluid helium experiments. Moreover, our study indicates that particles can be a useful tool for future studies of vortex motion in Bose-Einstein condensates (BECs). 

The advantage of our analytical and semi-analytical models is that they can be applied, and work better, for setups with the particles much larger than the healing length, whereas it would be virtually impossible to simulate such systems with huge scale separations in DNS. Such models will be useful, for example, in investigations of the effect of hydrogen ice tracers whose typical diameter is the order of a few micro-meters \cite{GBewley:2006aa}, while the vortex core diameter is of the order of an angstrom ($\AA = 10^{-10}m$). In this situation, we can think of a classical flow with a thin boundary layer so that classical textbook solutions relating the force on a moving object with the circulation around the object can be used \cite{acheson1990elementary}. This indicates that the Magnus model will work better in describing the motion of large particles relative to vortex core size. However, the validity of such a model in quantum fluids can not be taken for granted due to the presence of the compressibility, acceleration as well as a flow non-uniformity over the distances comparable to the particle sizes.

In our paper, we demonstrate that the Magnus force model works very well even for situations well beyond the formal limits of applicability of the ideal flow descriptions, e.g. when the size of the particle is no so much bigger than the healing length, and the velocity is not so much smaller than the speed of sound.

\section{Model and Numerical Methods}

\subsection{Gross-Pitaevskii equation coupled with particles}

We use the GP theory to model the superfluid flow and study its interaction with particles. The GP framework provides a good hydrodynamical description of a weakly interacting superfluid at low-temperatures and is able to reproduce the qualitative features of the strongly interacting superfluid helium. Within this framework, the state of the system is specified by the complex scalar field $\psi(\mathbf{r},t)$. 

The particles that we consider are \textit{active}; they are affected by flow and act back on it too. In our earlier works\cite{shuklaPRA16,pandit2017pofoverview,Vish2017} we introduced a Lagrangian for this combined system, wherein particles were represented by specifying the potential $V_{\mathcal{P}}$. This procedure yields the following GP equation for the spatiotemporal evolution of $\psi(\mathbf{r})$:
\begin{equation}\label{eq:GPEfull}
i\hbar\frac{\partial \psi}{\partial t} = -\frac{\hbar^2}{2m_b}\nabla^2\psi -\mu\psi + g|\psi|^2\psi
+ \sum^{\mathcal{N}_0}_{j=1}V_{\mathcal{P}}(\mathbf{r}-\mathbf{q}_j)\psi,
\end{equation}
where $\hbar = h/2\pi$ is the reduced Planck's constant, $m_b$ the mass of bosons constituting the superfluid, $g$ the effective interaction strength among these bosons, $\mu$ the chemical potential and $\mathbf{q}_j$ the position of the $j$th particle (i.e. center of the potential).

Our modelling of the particles by specifying $V_{\mathcal{P}}$ allows us to control their characteristics, e.g. shape and size. 
For the purpose of present study, we use the Gaussian potential
\begin{equation}\label{eq:potparticle}
V_{\mathcal{P}} = V_0\exp\Bigl(-\frac{r^2}{2d^2_p}\Bigr);
\end{equation}
here $V_0$ is the strength of the potential and $d_{\rm p}$ is the measure of its width.

Furthermore, we include a two-particle, short-range repulsion potential. Thus, the Newtonian dynamics of the particles is governed by the following equation,
\begin{equation}\label{eq:eqmpartfull}
	m_{j}\ddot{\mathbf{q}}_j = \mathbf{f}_{0,j} + \mathbf{G}_{j},
\end{equation}
where $m_j$ is the mass of the $j$th particle, which we assume to be same for all, $m_j=m_{0}$, vector $\mathbf{f}_{0,j}$ is the force exerted by the superfluid onto the particle,
\begin{equation}\label{eq:frcbyfluid}
	\mathbf{f}_{j} = \int_{\mathcal{A}}|\psi|^2\nabla V_{\mathcal{P}}(\mathbf{r}-\mathbf{q}_j)\,d\mathbf{r},
\end{equation}
${\mathcal{A}}$ is the area occupied by the particle (determined by a cutoff of the potential (\ref{eq:potparticle})), 
and $\mathbf{G}_{j}=(G_x,G_y)$ is the inter-particle short-range repulsion force.

\subsection{Numerical methods, units and parameters}

To study the dynamics of particles in complex superfluid flows, we solve Eqs.~\eqref{eq:GPEfull} and \eqref{eq:eqmpartfull} numerically. In order to do so, we perform direct numerical simulations (DNSs) of the GP by using the Fourier pseudospectral method on a square, periodic simulation domain $\mathcal{A}$ of side $L$ with $N^2_c$ collocation points~\cite{vmrnjp13}. In this method, we evaluate the linear terms in Fourier space and the nonlinear term in real (physical) space, which we then transform to Fourier space. For the Fourier-transform operations, we use the FFTW library~\cite{fftwsite}. A fourth-order, Runge-Kutta scheme is used to evolve these equations in time. Further details can be found in \cite{Vish2017}.

We define the length scale $\xi=\hbar/\sqrt{m_bg\rho_0}$, known as the healing length and the speed of sound $c=\sqrt{g\rho_0/m_b}$. The mean density can be also calculated $\rho_{\rm 0}=\int_{\mathcal{A}}\,|\psi|^2\,d\mathbf{r}/\mathcal{A}$. We then choose to rescale the parameters in \eqref{eq:GPEfull} with the following scalings: $\tilde{\mu}=\mu/(g\rho_0)$, $\tilde{V}_\mathcal{P}=V_\mathcal{P}/(g\rho_0)$, $\tilde{\psi}=\psi/\sqrt{\rho_0}$, $\tilde{\mathbf{r}}=\mathbf{r}/\xi$ and $\tilde{t}=tc/\xi$, we arrive at the dimensionless equation which we simulate:

\begin{equation}
i\frac{\partial \tilde{\psi}}{\partial \tilde{t}} = -\frac{1}{2}\tilde{\nabla}^2\tilde{\psi} -\tilde{\mu}\tilde{\psi} + |\tilde{\psi}|^2\tilde{\psi}
+ \sum^{\mathcal{N}_0}_{j=1}\tilde{V}_{\mathcal{P}}(\mathbf{r}-\mathbf{q}_j)\psi.\label{eq:GPEsim}
\end{equation}
 From here on we drop the tildes for simplicity of notation. Taking into account that we are in dimensionless units, in our calculations we have $\rho_{\rm 0}=c=\xi=\mu=1$. In dimensionless units we also choose the parameters $L=177.78$, with the grid spacing $dx=L/N_c$, and the number of collocation points $N_c=256$. For the external potential we choose the parameters $V_{\rm 0}=10$, and $d_{\rm p}=1.5$. To calculate the initial conditions we solve the real Ginzburg-Landau equation (RGLE), this minimises the energy of the fluid in the presence of the potentials modelling the particles. 
 
\subsection{Magnus force model}

Lift force or the Magnus effect is a well-studied phenomenon in classical fluid dynamics \cite{tritton1988physical}. In a fluid flow with a uniform upstream velocity $\mathbf{u}_{\rm flow}$, a cylindrical disk with circulation $\Gamma$ around it experiences a lift force $\rho \Gamma \mathbf{u}_{\rm flow}\times \mathbf{\hat{e}}_z $ , where $\rho$ is the fluid density and $\mathbf{\hat{e}}_z$ is the unit vector along the cylinder corresponding to the vorticity direction. 
This phenomenon in superfluid was first observed by Vinen \cite{Vinen:1961aa} by measuring modifications to frequency of a vibrating wire submerged in He II, which also allowed to demonstrate the quantisation of the circulation around the wire.
In the present work, the fluid flow relative to a solid object is induced not
by mechanical properties of the solid object itself (e.g., its elasticity), but rather
by an external vortex (or multiple vortices) not trapped by this particular object.

Here, we want to explore the dynamics of an assembly of particles trapped on 2D vortices in superfluids. In particular, we want to elucidate the role of the Magnus force acting on these particles. To this end, we develop a Magnus force model (MFM) to describe this system.
The Magnus force induced acceleration of the $j$th particle trapped on a vortex (circulation strength $\Gamma$) at the location $(x_j,y_j)$ is
given by
\begin{align}
\ddot{x}_j &= A(\dot{y}_j- {v}_j) + G_{x}/(m_0+m') \label{2Dx},\\
\ddot{y}_j &= A({u}_j-\dot{x}_j) + G_{y}/(m_0+m') \label{2Dy},
\end{align}
where the overhead dots indicate time derivative; $({u}_j,{v}_j)$ is the flow velocity around the particle collectively induced by the other (excluding the $j$th) vortices. In our description of the particle dynamics, we ignore the variation of the flow velocity over a distance comparable to the particle radius. Such an approximation is valid when the distance between the particles is large compared to both the particle radius and the healing length. The parameter $A=\Gamma\rho/(m_0+ m')$ is the natural oscillation frequency of the vortex trapped particle in our system. $m_0$ is the physical mass of the particle and $m'$ the hydrodynamical added masses (see below).

Note that in \eqref{2Dx} and \eqref{2Dy} we have introduced a short-range repulsion $G$ between the particles, which acts only when the particles come to distances comparable to their size. Strictly speaking the approximation that the velocity around a particle is uniform fails at such short distances; however, our goal in this paper is to test the model beyond its formal limits of applicability in order to test its robustness.
The added mass can be computed using the unsteady Bernoulli equation (derived from the GPE),
 \begin{align}\label{BernoulliEq}
 \frac{\partial \phi}{\partial t} +\frac{1}{2}(\nabla \phi)^2 -\sum_{j=1}^{\mathcal{N}_0} V_{\mathcal{P}}(\mathbf{r}-\mathbf{q}_j) = \frac{1}{2}\frac{\nabla^2 \sqrt{\rho}}{\sqrt{\rho}} - \frac{p}{\rho},
 \end{align}
where $ \phi = \arg \psi$, which relates to the superfluid velocity $\mathbf{v}$ via $\mathbf{v}=\nabla \phi$,
density $\rho = |\psi|^2$ and the first term on the right hand side in Eq.~\eqref{BernoulliEq} is the quantum pressure term. Note that in this model the dynamics are compressible and that $\rho=\rho(\mathbf{x},t)$.
 
 Let us assume for the moment that the potential representing the particle is hard, i.e., it has a well defined boundary (an extension to the case of ``soft'' particle potential is made later). In such a case, the superfluid density $\rho$ is zero within the particle boundary and it ``heals'' to its bulk value over a boundary layer that is approximately healing length wide. Now, if the particle radius $R \gg \xi$, then we can regard the particle and its boundary layer as a single moving control volume. This allows us to neglect the quantum pressure at the boundary of the considered control volume, and we can write
 \begin{align}
p = -\rho \left(\frac{\partial \phi}{\partial t} + \frac{1}{2}|\nabla \phi|^2\right),
\end{align}
i.e. the classical expression for irrotational ideal fluids. Thus, we can apply the classical textbook calculations for both the Magnus force and the added mass by integrating the pressure distribution over the control volume boundary\cite{lamb1916}. If the particle under consideration is a 2D disk of radius $R$, then the $\frac{\partial \phi}{\partial t}$ term gives rise to the added mass $m'=\rho \pi R^2$. 

Our method of treating the dynamics of particle loaded vortices self-consistently takes into account the variations in the background flow velocity around the particles because of the dynamically varying separation between the vortices. It is important to appreciate the fact that Magnus force description given by Eqs.~\eqref{2Dx} and \eqref{2Dy} depends on the motion of the particle relative to the fluid and the resulting force is perpendicular to the motion. In the other words, there would be no Magnus force if the particle motion was tracing the fluid paths (this regime would be realised by the limit of very small and light particles).
 
\begin{figure}[h!]
	\center{
		\includegraphics[width=\linewidth]{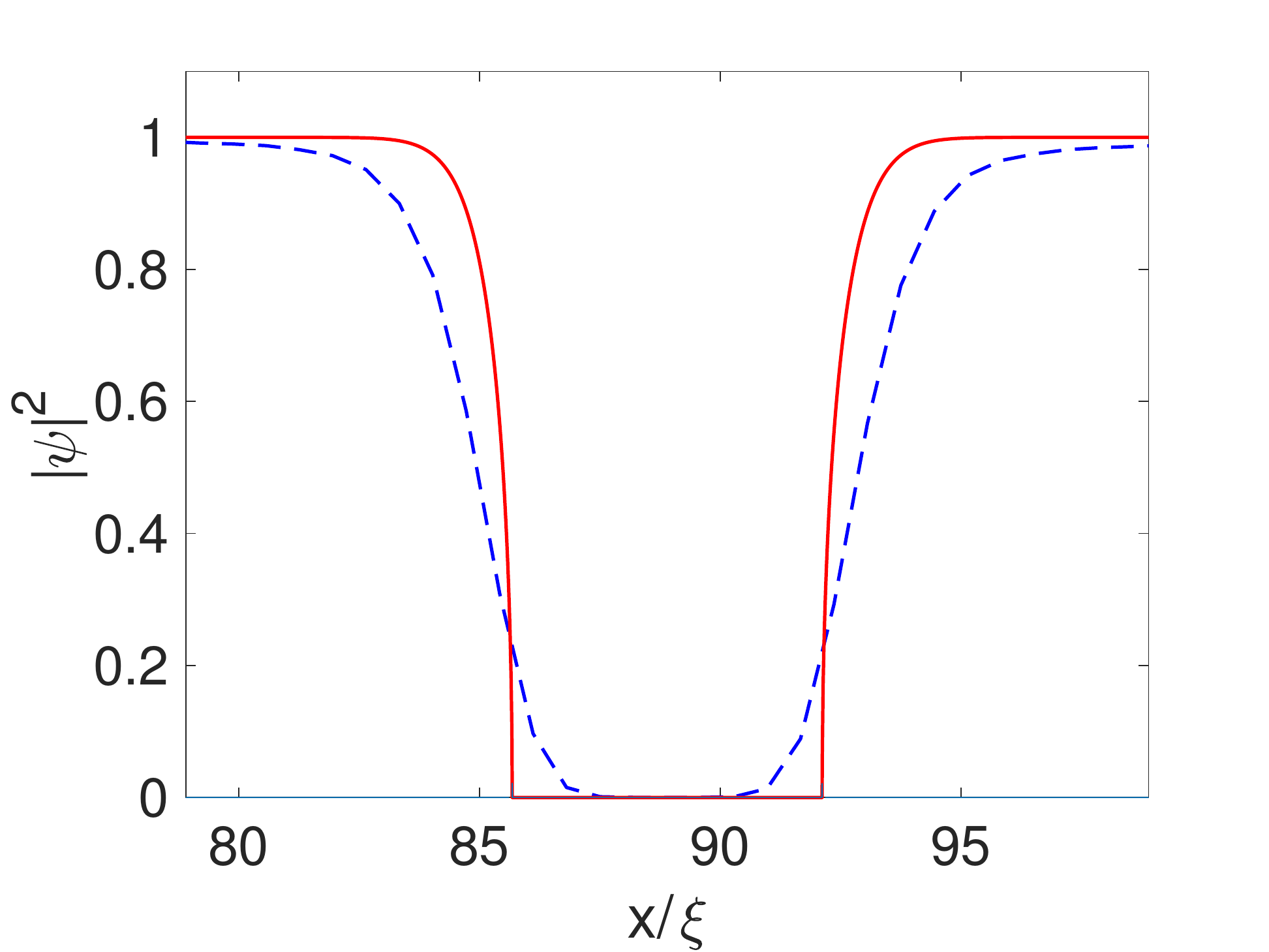}
		\caption{{\it (Color online)} Density profile of particle and TF profile: The density profile of the simulated particle is blue (dashed), The TF profile is red, both with the same parameters, $V_0=10g\rho_0$ and $d_p=1.5\xi$.}
		\label{TF}}
\end{figure}

In our GP simulations, we use a Gaussian potential, which has a soft boundary, to represent particles; therefore, we need an estimate of an effective radius to compute the added mass $m'$. A consequence of the soft boundary is that the fluid can penetrate a region of the potential, but is slowed down by the increasing strength of the potential until the reflection point. A correct estimation of the effective radius is a non-trivial exercise. Therefore, we compute it in an ad-hoc manner by making use of the Thomas-Fermi (TF) profile of the superfluid density around the particle potential. 
The TF profile is the density of a time-independent solution of Eq.~\eqref{eq:GPEsim} in which we remove the Laplacian term, this approximates the density depletion due to the particle on the surrounding fluid with no kinetic energy.
The TF profile $|\psi_{TF}|^2$ in the presence of a single particle is then given by
\begin{align}
|\psi_{TF}|^2=\left(1-V_0e^{\frac{-r^2}{2d_p^2}}\right)\mathcal{H}(r-R_{TF}),\label{TF1}
\end{align}
where $\mathcal{H}$ denotes the Heavyside step function and $R_{TF}$ the radius of the region within which the profile is zero. 
In Fig.~\ref{TF} we show a slice of the TF profile, along with the initial condition. The initial condition is a slice across the particle as calculated by solving the RGLE with the potentials modelling the particles.
The radius $R_{TF}$, computed from the Eq.~\eqref{TF1}, is given by
\begin{align}
R_{TF}=d_p\sqrt{2\log(V_0)}.
\end{align}
The TF profile allows us to obtain an estimate of the displaced mass of the superfluid due to the particle (while fluid density is held fixed at $\rho_0$). We use this displaced mass as the added mass, as this captures the features of the soft potential. Thus,

\begin{align}
m'=2\pi\int_{0}^{\infty} (1-|\psi_{TF}|^2) r dr.
\end{align}

To be more clear here, the added mass can be decomposed into two parts: $m'_1$, the contribution governed by the geometry of the particle, while assuming incompressibility; $m'_2$, the contribution coming from the boundary layer. We express this as follows:
\begin{align}
m'=m'_1 + m'_2 & =\pi R_{TF}^2 + 2\pi\int_{R_{TF}}^{\infty} V_0e^{-\frac{r^2}{2d_p^2}} r dr\\
&= \pi R_{TF}^2 + V_0d_p^2e^{-\frac{R_{TF}^2}{2d_p^2}} \nonumber \\
&= d_p^2(2\pi \log(V_0) + 1).
\end{align}
It is worth emphasising that the added mass depends on the geometry of the particle, and in the case of an arbitrarily shaped particles further considerations are required compared to what we wish to address in the present study. For the parameters used in this study, we find $m'$ to be $46.69$.

 We use the added mass $m'$ of the displaced superfluid from the TF profile to define the ratio
\begin{equation}
	\mathcal{M}\equiv \frac{m_0}{m'},
\end{equation}
which allows us to distinguish between heavy ($\mathcal{M}>1$), neutral
($\mathcal{M}=1$) and light ($\mathcal{M}<1$) particles. We emphasise here that the added mass in general differers from the displaced mass; however, for disk shaped particles they are identical.


We express the MFM equations in a more compact form as
\begin{align}
\ddot{z}_j &= iA(w_j-\dot{z}_j) + G/(m_0+m') \label{2Dz} ,
\end{align}
where $z_j=x_j+iy_j$, $w_j=u_j+iv_j$ and $G=G_x +iG_y$. 

We can easily extend our 2D MFM to 3D by using the Local Induction Approximation (LIA) of the Biot-Savart law, 
wherein a vortex element $\mathbf{s}$ of a vortex line at the arc length $\zeta$ and time $t$ has velocity\cite{Schwarz1988}
\begin{align}
\dot{\mathbf{s}}(\zeta,t) &=\beta \mathbf{s}' \times \mathbf{s}'', \label{LIA} 
\end{align}
where the overhead dot and prime correspond to derivatives with respect to the time and arclength, respectively; $\beta=\log(l/\xi)$ and $l$ is a suitable cut-off length scale that is approximately equal to the mean curvature radius.
 
Therefore, in 3D the dynamics of the vortex-line, with uniform mass distribution, is given by 
{
\begin{align}
\ddot{\mathbf{q}} &= A\mathbf{s}'\times ( \dot{\mathbf{q}} - \dot{\mathbf{s}}(\zeta,t)),\label{3DMod} 
\end{align}}
where $\mathbf{q}(\zeta,t)$ is the position of the particle. This equation corresponds to a vortex line with test particles densely filling its core. Another interpretation could be normal fluid trapped into the superfluid vortex core.
Similar to the 2D case, in 3D we can include a short range repulsion between the particles.

The LIA is derived from the more general Biot-Savart law. Using the Biot-Savart description with our model requires more complexity. This is due to the vortex arclength not being conserved in the Biot-Savart description. Thus, a separate equation for the mass density along the vortex line would be required.

\begin{figure}[h!]
	\center{
		\includegraphics[width=\linewidth]{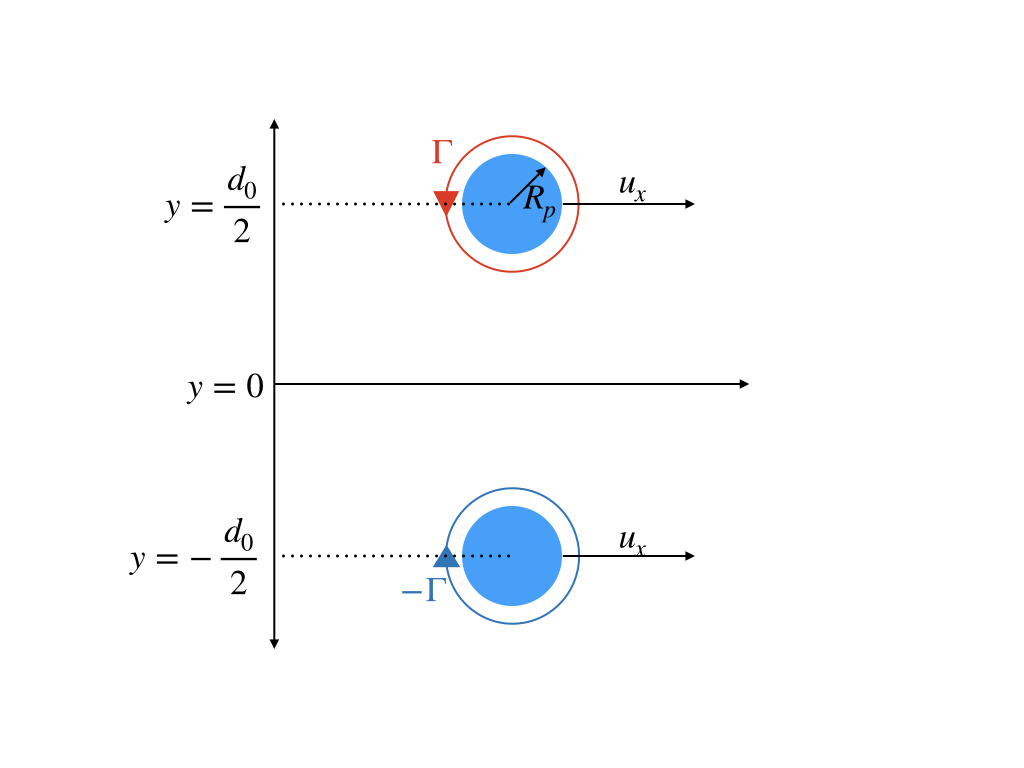}
		\caption{{\it(Color online)} Schematic diagram illustrates the initial configuration, wherein a vortex-antivortex pair loaded with particles is translating along the x-axis. Particles are shown by blue disks. The circle with anticlockwise (clockwise) arrow represents a vortex (an antivortex) with circulation vector pointing out of (in) the plane. The centers of the particles (the coincident vortices) are separated by a distance $d_0$.
%
}\label{Schem}}
\end{figure}

\subsection{Simplified Magnus force model for particle loaded vortex-antivortex pair}

A vortex-antivortex pair of size $d_0$ is the simplest multi-vortex configuration that occurs in a periodic, 2D domain. This vortex-antivortex pair translates at a speed $u=\Gamma/(2\pi d_0)$ in a direction perpendicular to the line joining the two vortices (See Fig.~\ref{Schem}). Therefore, the dynamics of the two particles $\mathcal{P}_1$ and $\mathcal{P}_2$ trapped on the vortex and antivortex, respectively, serves to provide a simple demonstration of the Magnus effect. 

To simplify our discussion, we assume that the $y$-component of the velocity of the underlying flow experienced by the particles is zero, i.e. $v=0$
 in Eqs.~\eqref{2Dx} and \eqref{2Dy}. Also, we consider a large vortex-antivortex pair, $d_0 \gg R_p$, that allows us to neglect the short-range repulsion. As our vortex-antivortex pair is symmetric about the $x$-axis (See Fig.~\ref{Schem}), in what follows we discuss the dynamics (trajectory) of only one particle. The equation of motion for the particle is then given by
 \begin{align}
 \ddot{x} &= -A\dot{y}, \label{simplex}\\
 \ddot{y} &= -A(u(y) - \dot{x}). \label{simpley}
 \end{align} 
 To further simplify the discussion, we impose a condition that the horizontal component $u(y)=\text{constant}$ during the dynamical evolution of this system, thereby furnishing a readily solvable set of coupled ODEs. Hereafter, we refer to this model as the simple Magnus force model (SMFM). In SMFM, with the initial conditions 
 \begin{align}
 x(0)=x_0 && y(0)=y_0 \\
 \dot{x}(0) = 0 && \dot{y}(0)=0, \nonumber
 \end{align}
 the particle trajectory is of the following form:
 \begin{align}
 x(t) &=x_0 +ut-\frac{u}{A}\sin(At),\label{uconstx} \\
 y(t) &=y_0 -\frac{u}{A}(1-\cos(At)). \label{uconsty}
 \end{align}
 Note that this rather restrictive description is valid only when the oscillations of the particles are small. 
 
 In the present study, we compare the predictions of the MFM and SMFM against the GP description to illustrate the Magnus effect. Therefore, it is important to recognise the fact that due to the periodicity of the phase of the wave function representing the vortex-antivortex pair within the GP description in a 2D periodic domain, the motion of the pair is altered as compared to that in the ideal fluid. We discuss this in detail in appendix \ref{DV}, where we provide a detailed comparison of the vortex-antivortex pair dynamics in the GPE and the ideal fluid case (Weiss-McWilliams formula).

\section{Results}
\subsection{Dipole configuration}

\begin{figure}
	\center{
		\includegraphics[width=0.95\linewidth]{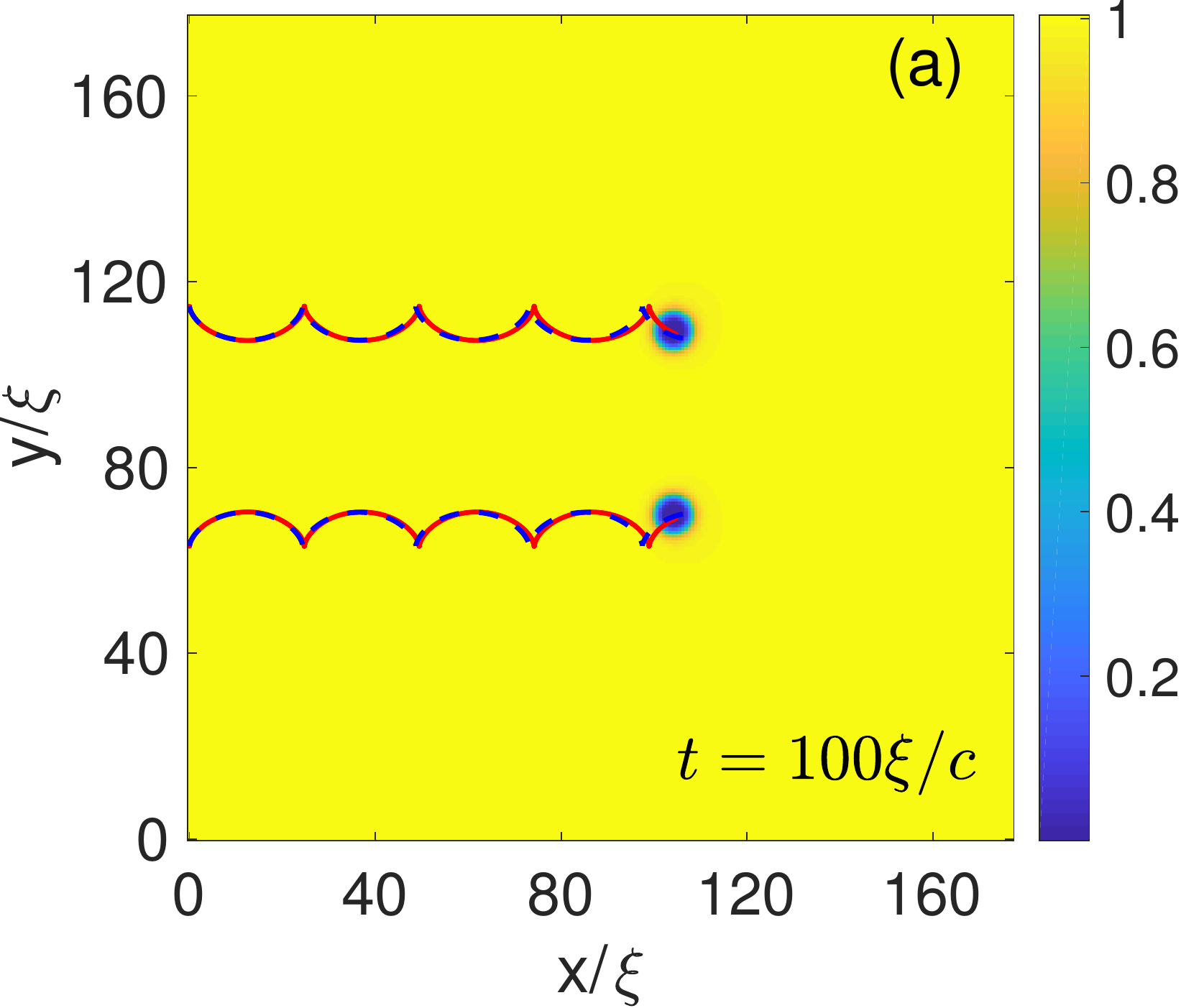}
		\caption{{\it (Color online)} Trajectories of loaded vortex dipole. Parameters: mass ratio $\mathcal{M}=17.15$ and initial separation $d_0= 52.50 \xi$. Pseudo-color density ($|\psi(\mathbf{r},t_f)|^2$) plots with simulated (red) and predicted with MFM (blue dashed) trajectories overlaid.}
		\label{M1P029_dens}}
\end{figure}

We first discuss the discuss the dynamics of the particle loaded vortex-antivortex pair, starting with an initial configuration as shown in Fig.~\ref{Schem}. In Fig.~\ref{M1P029_dens}, we show the pseudo-color plots of the density field $\rho(\mathbf{r})$ overlaid with trajectories of the two particles, along with SMFM predictions (blue dashed curves) given by Eq.\eqref{uconstx} and Eq.\eqref{uconsty}. We find that the particles follow a nearly cycloid trajectory, in good agreement with the SMFM predictions. The cycloid trajectory of the particles is characterised by the displacement amplitude in the $y$-direction $\delta_a$ and the periodic length $X_p$ in the $x$-direction. These two quantities are easily deduced from the SMFM (Eqs.~\eqref{uconstx} and \eqref{uconsty} ) yielding $\delta_a=2u/A$ and $X_p\simeq u 2\pi/A$; note the dependence on the flow velocity $u$.



\begin{figure*}
\center{
\includegraphics[width=0.49\linewidth]{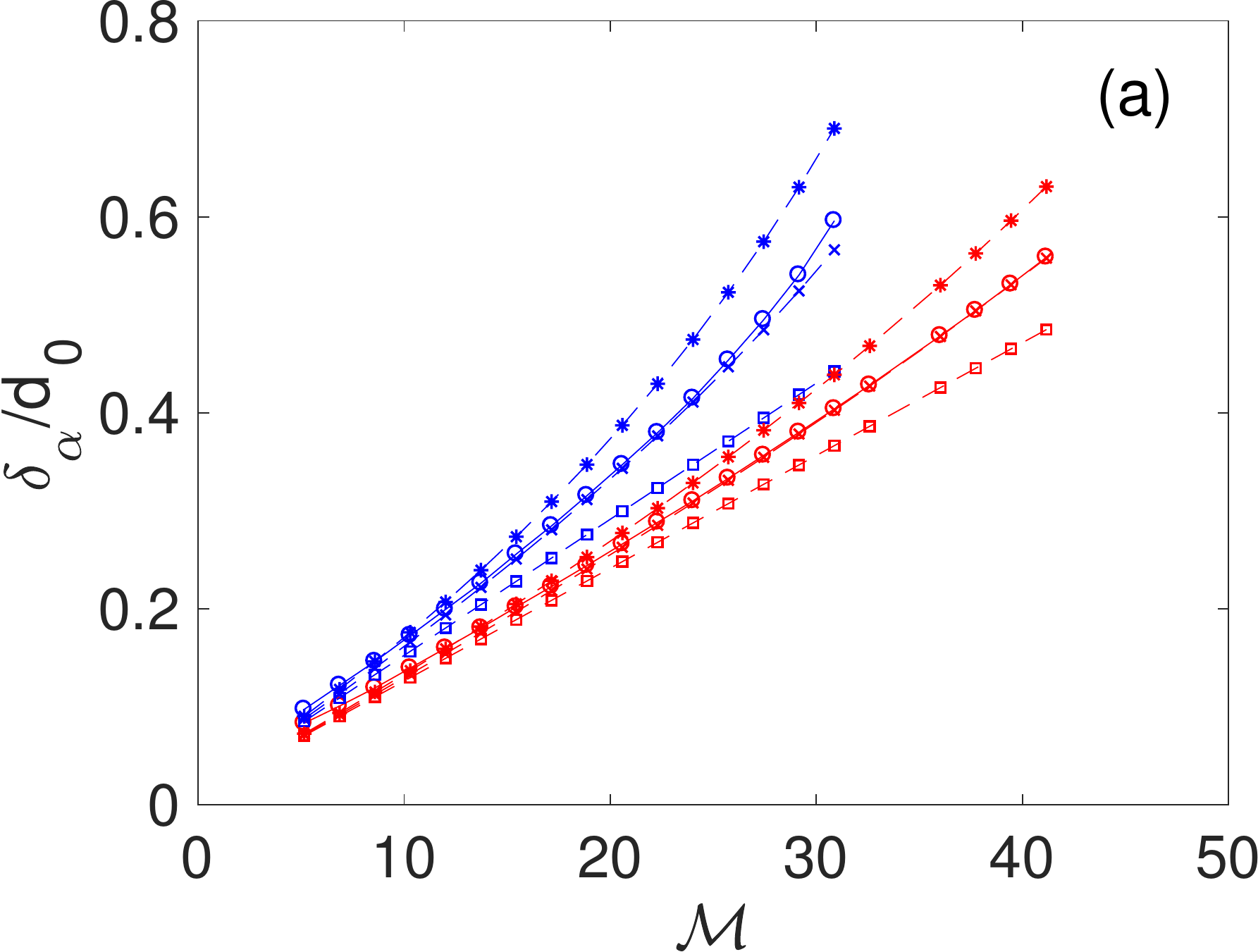}
 \includegraphics[width=0.49\linewidth]{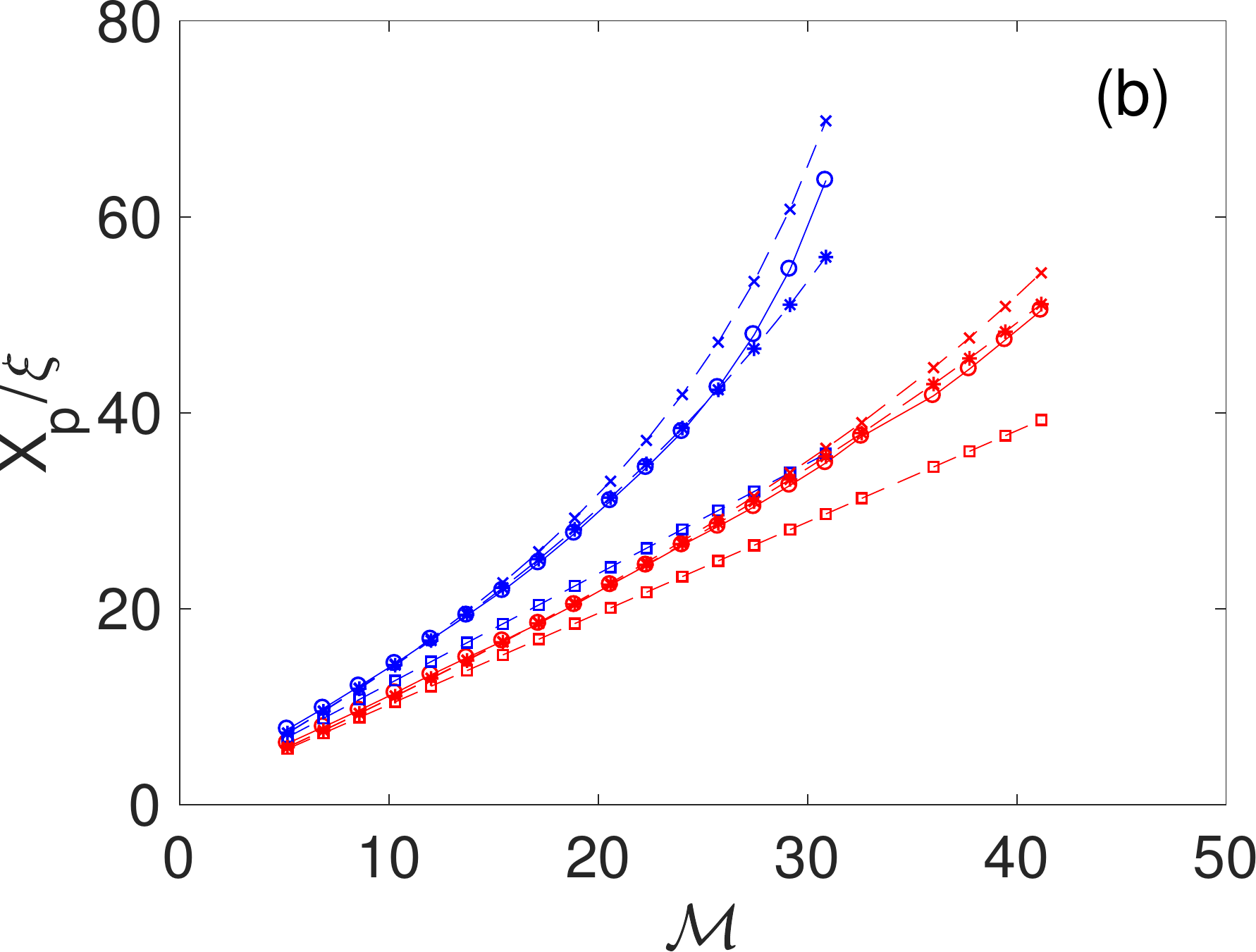}
\caption{{\it (Color online)} Amplitude and period analysis of SMFM. All blue (top four) lines correspond to a separation $d_0= 52.50 \xi$ and red lines (bottom four) to $d_0= 72.41\xi$ with the circles representing simulated data in both (a,b). The other types of line correspond to the fit for a velocity $u_{pair}(d_0)$ (square, case 1 in text), for velocity $u_{pair}(d_0-2\delta_\alpha)$ (asterisk, case 2 in text) and average of the other two (cross, case 3 in text). Panel (a): the amplitude $\delta_\alpha$ of the cycloid in units of initial separation against mass ratio $\mathcal{M}$. Panel (b): the period $X_p$ against mass ratio $\mathcal{M}$.\label{AMP}}}
\end{figure*}

To better appreciate this dependence, we approximate the flow velocity $u$ by the following three values, with $u_{pair}$ given by Eq.~\eqref{AMCW}. 
\begin{enumerate}
\item $u= u_{pair}(d_0)$;
\item $u= u_{pair}(d_0 - 2\delta_0)$, where $\delta_0=2u_{pair}(d_0)/A$ is the amplitude of case (1); 
\item $u$ is the mean of the estimates obtained from (1) and (2) above.
\end{enumerate}
In Fig.~\ref{AMP} (a) and (b) we show the plots of $\delta_a/d_0$ and $X_p/\xi$, respectively, vs. $\mathcal{M}$ obtained from the GPE simulations (curves with circles as markers) and the use of the above three test cases for $u$ in SMFM; we do this for two initial values of $d_0=52.30\xi$ (blue curves) and $d_0=72.41\xi$ (red lines). This exercise reveals a clear dependence of the results on the choice of $u$; thereby suggesting that in any modelling scheme based on the MFM model, the flow velocity $u$ must be updated in a self-consistent manner.
Note that the case (1) is the velocity at the initial time, this also corresponds to the minimum velocity, as the particles are at maximum separation. Case (2), which tends to overshoot, is based on the approximate minimum distance, thus resulting in maximum velocity. Case (3) is a simple average, we see that this predicts well the parameters of the cycloid trajectories for all the simulations. 

\begin{figure*}[!t]
\center{
 \includegraphics[width=0.35\linewidth]{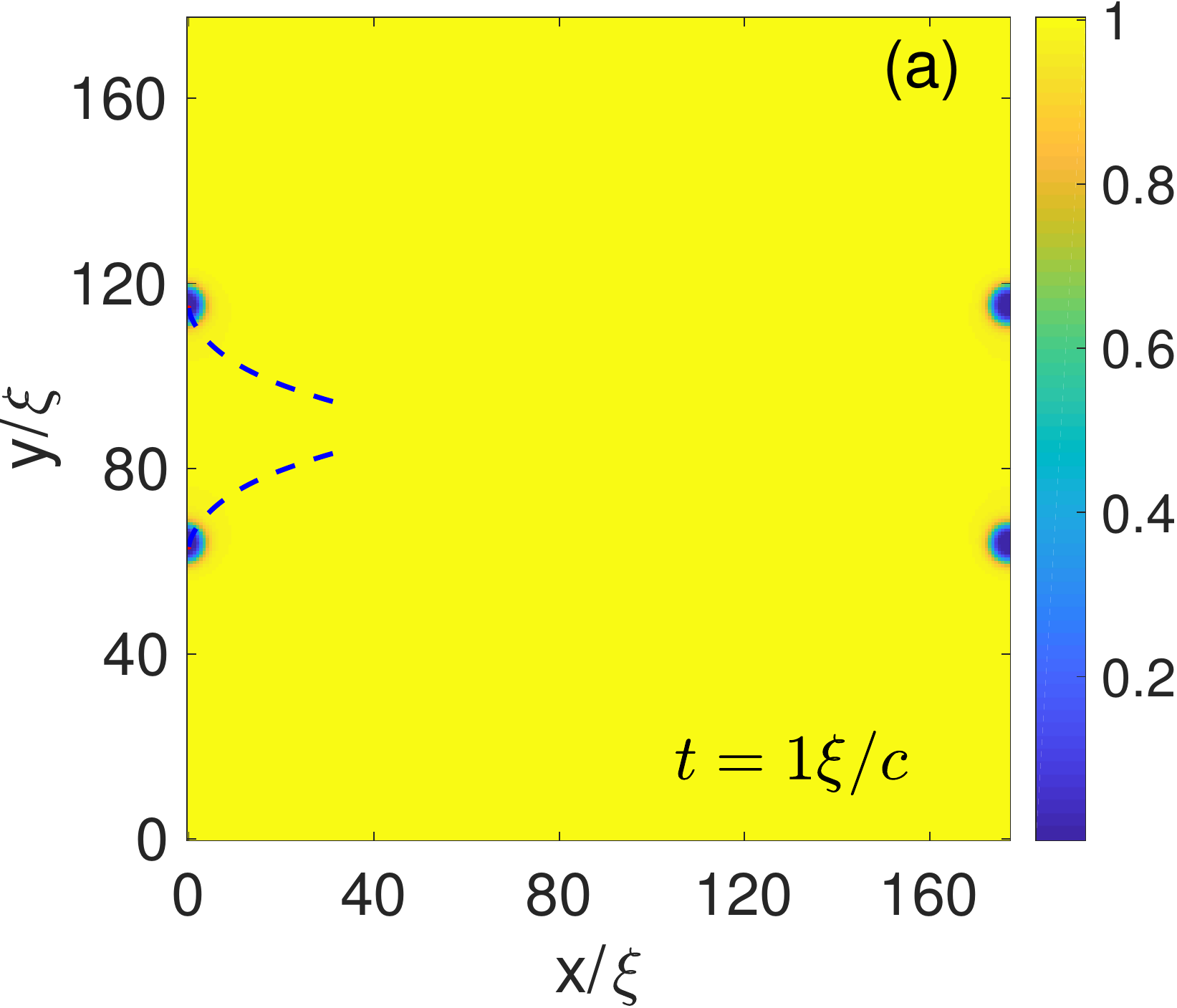} \hspace{6mm}
 \includegraphics[width=0.35\linewidth]{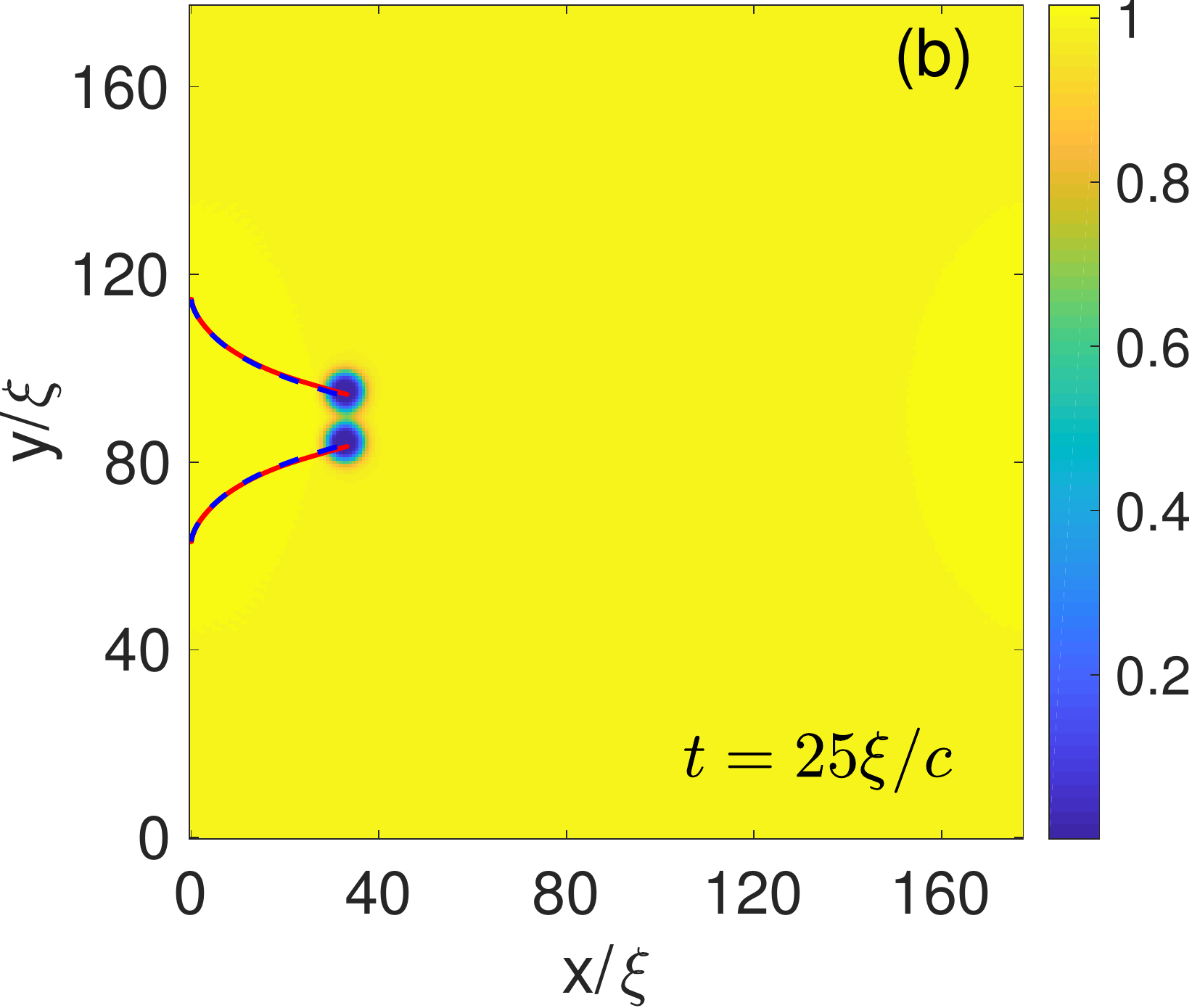}\\
 \includegraphics[width=0.35\linewidth]{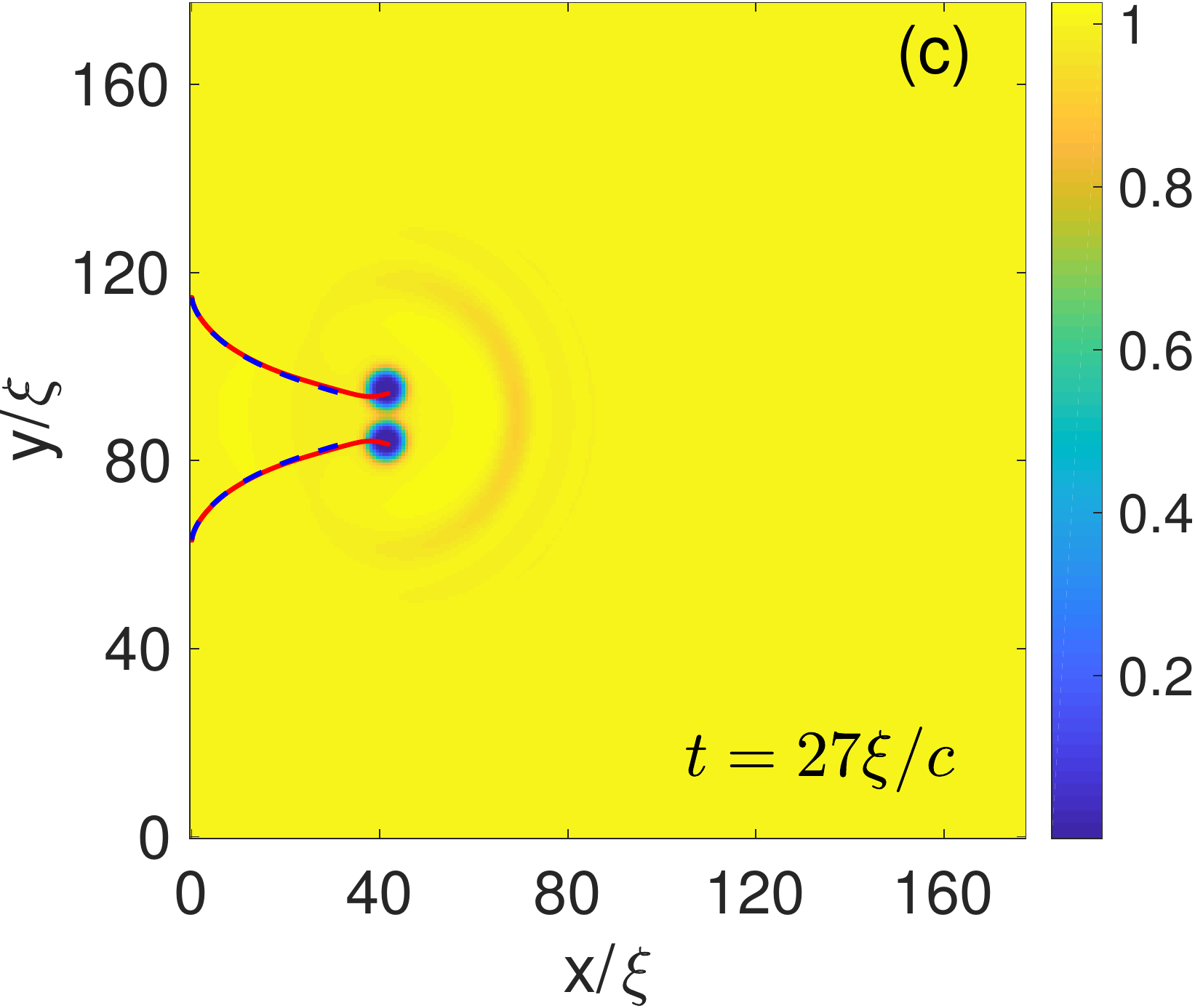} \hspace{6mm}
 \includegraphics[width=0.35\linewidth]{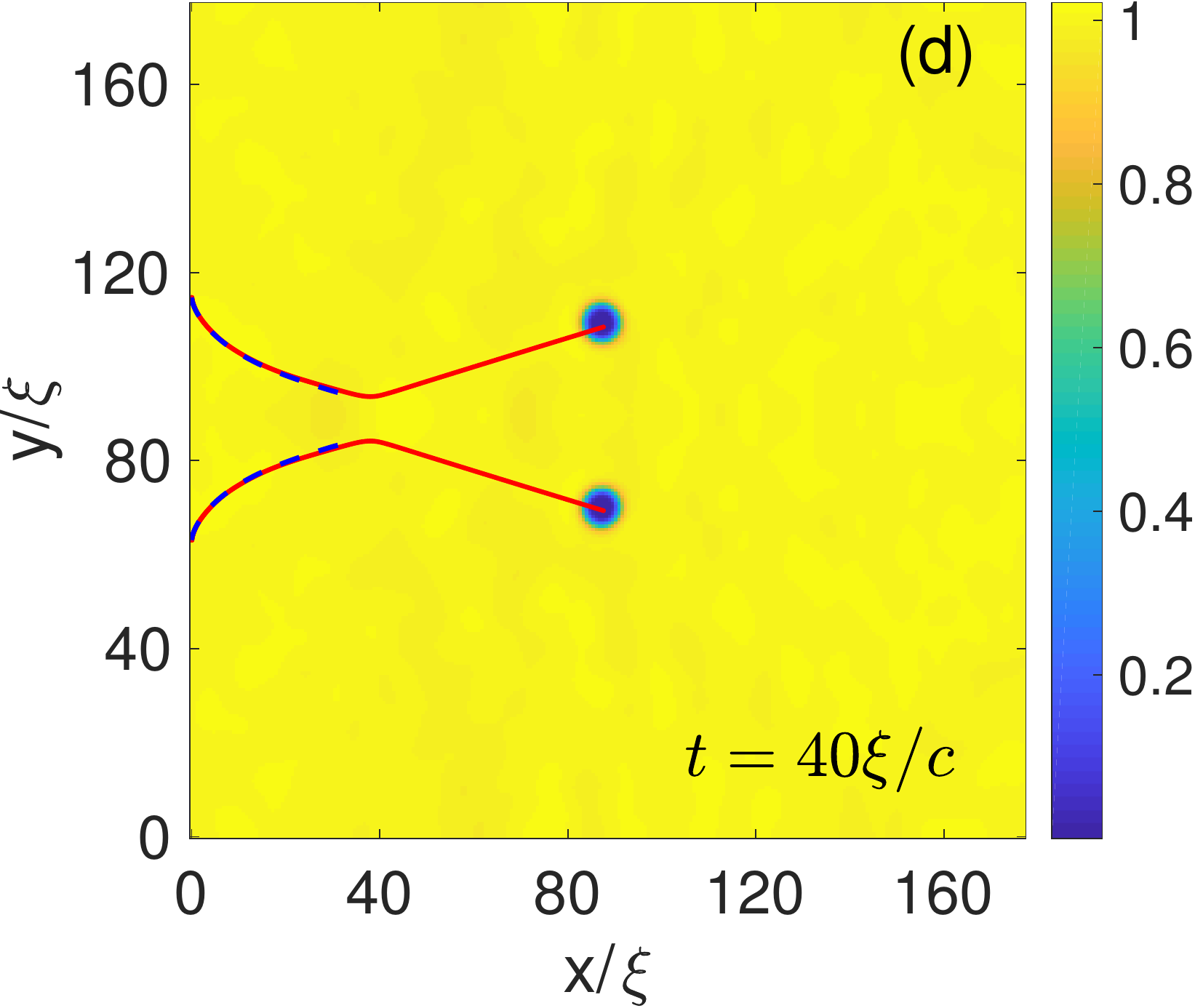}
\caption{{\it (Color online)} Collision of loaded vortices. Parameters: mass ratio $\mathcal{M}=36.02$ and separation $d_0= 52.5 \xi$. Psuedo-color density ($|\psi(\mathbf{r},t_f)|^2$) plots with DNS (red) and MFM (blue dashed) trajectories overlaid with time presented in the each panel. Sound produced during annihilation can be seen as variations in the psuedo-color density. The MFM trajectories are only presented up until the collision.}
\label{M21P29}}
\end{figure*}

Also note that in Fig.~\ref{AMP} the data for $d_0=52.50\,\xi$ and $\mathcal{M}>30.87$ is absent, as for these values of the parameters the amplitude of the cycloid motion becomes larger than half the initial vortex-antivortex pair size and results in a collision of the two particles, thereby annihilating the vortices. Later (in Fig.~\ref{M21P29}) we show that the model fits well up until the collision.

Now let us recall an assumption of our model, that the flow is uniform across the particle. This assumption is good as long as the vortex-antivortex pair size is large. However, as the size decreases, the finite size of the particles becomes increasingly important due to the external flow varying over the particle. Despite these restrictions, the analytic model provides a good phenomenological description of the dynamics of the particle loaded vortices.

To improve the simplified model we remove the assumption that the velocity is constant throughout the dynamics, namely, we let the velocity vary depending on the current separation of the particles. We solve the MFM Eqs.~\eqref{2Dx} and \eqref{2Dy} numerically, while accounting for the variation in $u\equiv u_{\rm pair}(d(t))$ as the vortex-antivortex pair size varies; this allows us to improve the accuracy. We compare the trajectories of the particles obtained directly from the GP simulations with those predicted by the MFM. We solve the coupled ODEs of the MFM by using the specialised ODE solver ODE45 in MATLAB.
We also include the short range repulsion forces as now they are relevant for the cases where our model is tested beyond the formal limit of applicability, namely, when the inequality $\xi \ll R \ll d$ does not hold.



For comparing the above two solutions, we define an average error
\begin{align}
\epsilon= \sum_{i=1}^{N_t}({(x_s(t_i) - x_p(t_i))^2 +(y_s(t_i) -y_p(t_i))^2})^{1/2}/N_t,\label{error}
\end{align}
where $(x_s,y_s)$ are the GP simulated coordinates of the particle and $(x_p,y_p)$ are the coordinates from the simulated ODE. We choose $N_t$ points on each trajectory which are evenly spaced in time. The measure of error, $\epsilon$, is the average distance the predicted value is away from the true value over the entire run, we present this in units of $\xi$.


\begin{figure}[h!]
\center{
 \includegraphics[width=0.95\linewidth]{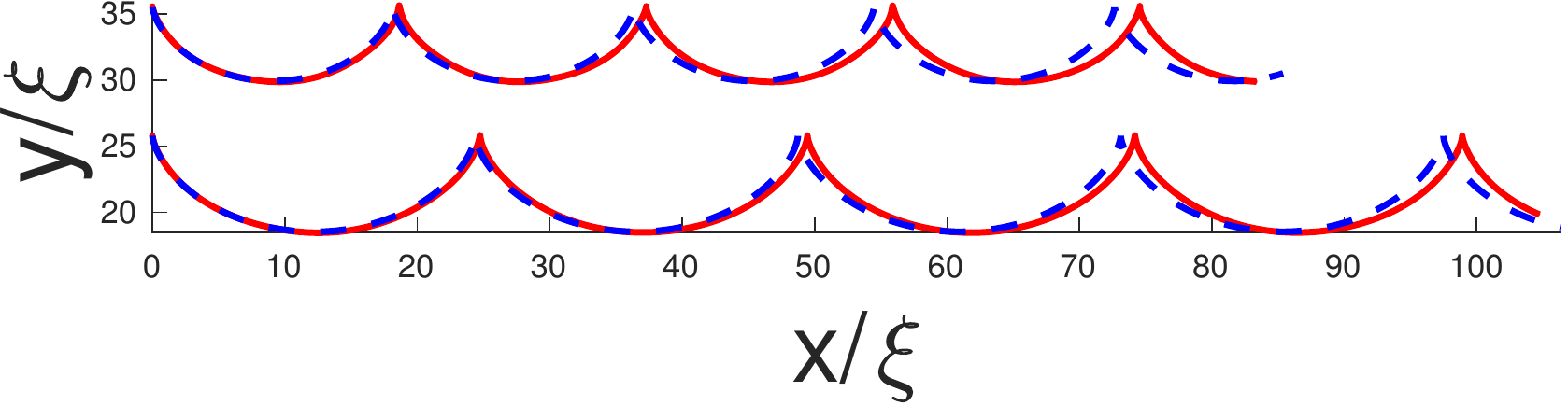}
\caption{{\it (Color online)} Trajectories of the loaded upper vortex in the dipole. Example of DNS trajectories (red) compared to predicted trajectories (blue dashed). Two cases show $d_0 = 52.50 \xi$ (bottom curve), $72.41\xi$ (top curve) with $\mathcal{M}=17.15$. The final time of the simulation is $t_f=100$.\label{M150_P2}}}
\end{figure}

In Fig.~\ref{M150_P2}, we present two examples ($d_0/\xi=52.50$ and $72.41$) of direct comparison between the particle trajectories obtained from the GP simulations and those predicted by the MFM. Note that we only show the trajectory of the particle trapped on a vortex with positive circulation because of the symmetry. The mismatch between the trajectories gives a visual indication of the error.

\begin{figure}
\center{
 \includegraphics[width=0.49\linewidth]{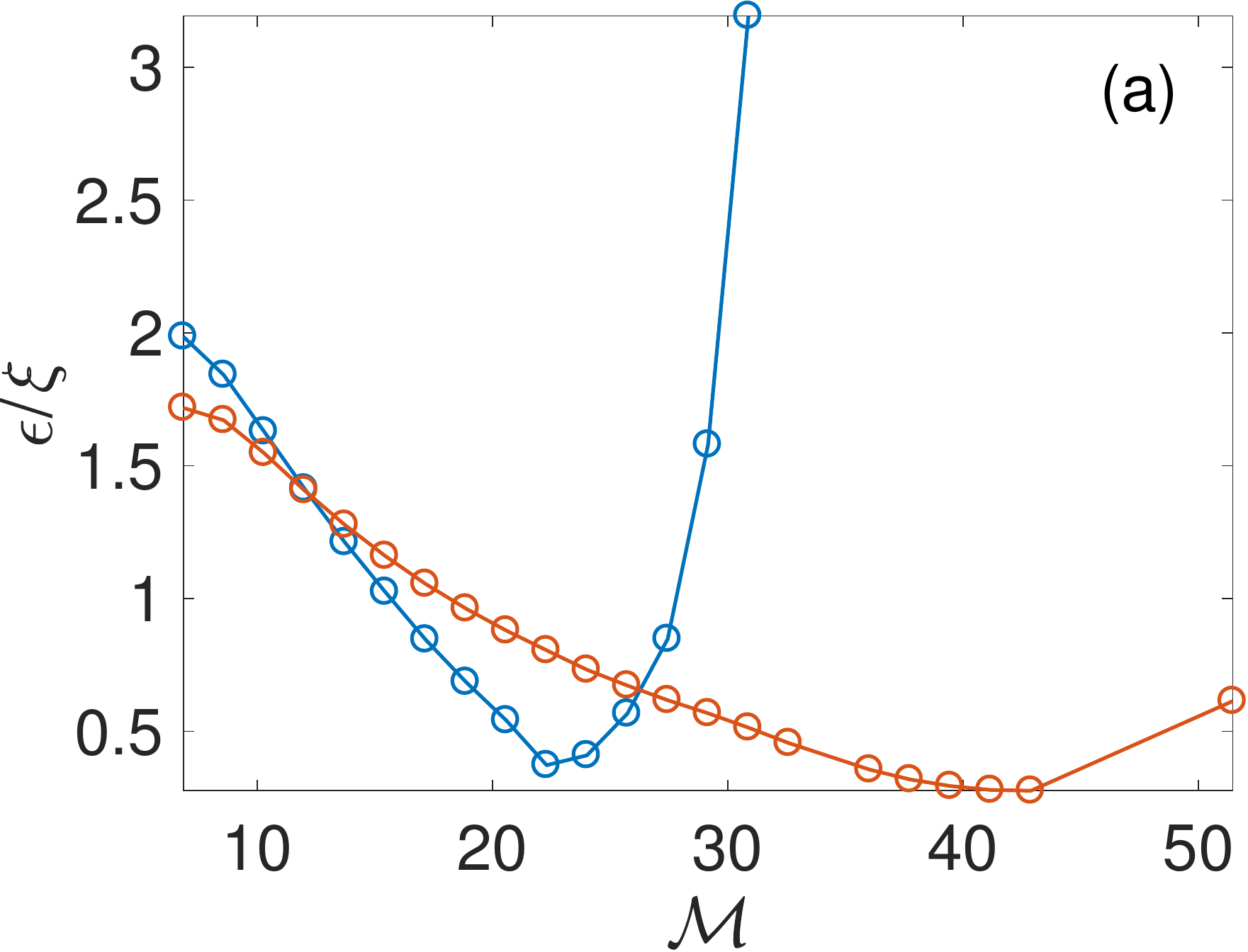}
 \includegraphics[width=0.49\linewidth]{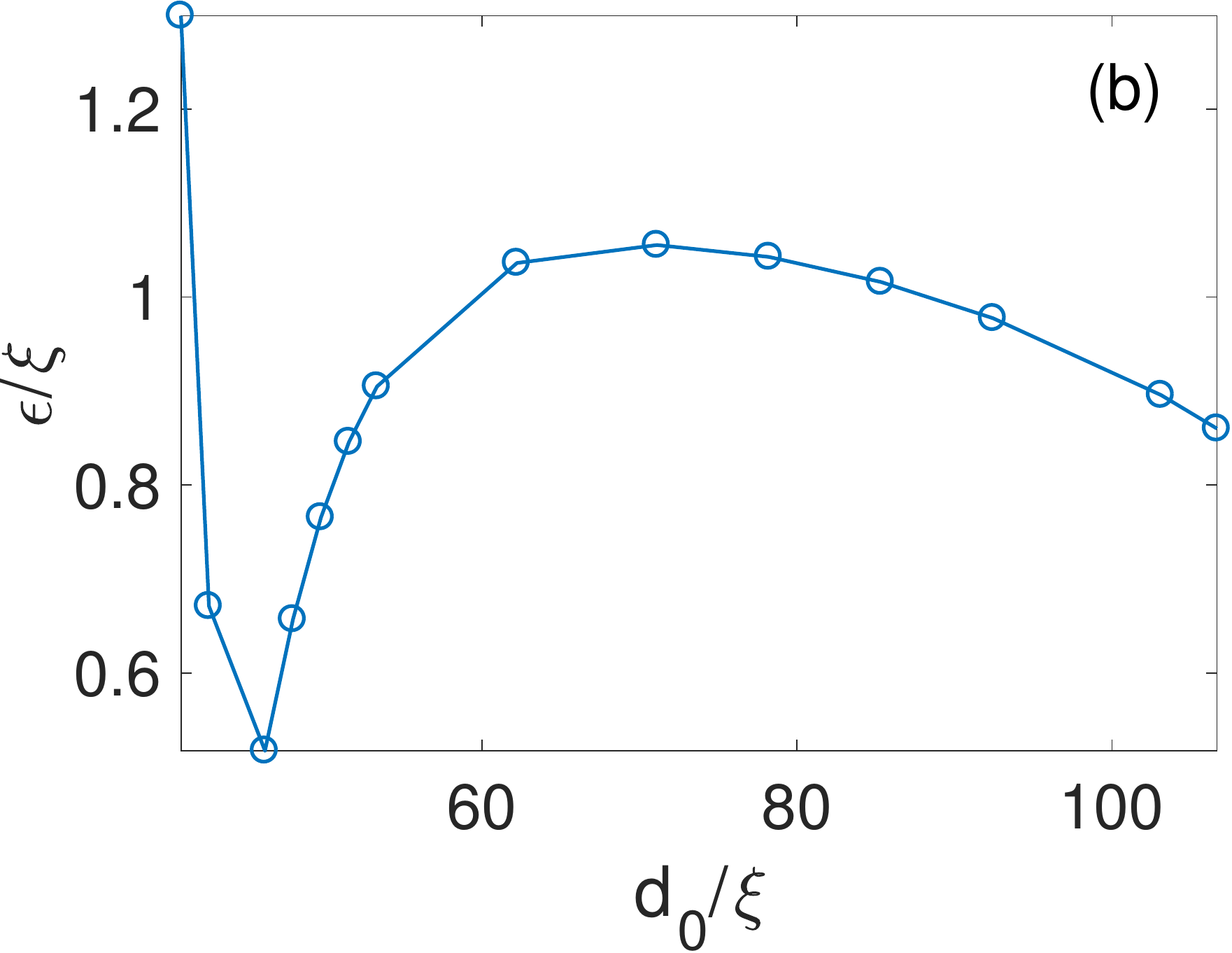}
\caption{{\it (Color online)} Error with simple dipole configuration. Panel (a): two cases show $d_0 = 52.50 \xi$ (blue) and $72.41\xi$ (red) for a range of different masses. For values of $\mathcal{M}$, greater than $30.87$ the particles have collided for $d_0 = 52.50 \xi$ (blue), so they have been omitted. Panel (b): For $\mathcal{M}=17.15$ with a range of different initial distances.}
\label{ERRM}}
\end{figure}

We perform numerical simulations for a large range of particle masses $\mathcal{M} \in [5,50]$ for the same two initial separations, keeping the total time of the dynamics fixed at $t_f = 100\xi/c$. We then compute the error $\epsilon$ and present the results in Fig.~\ref{ERRM} (a). We observe that the error for all simulations stays of the order of a healing length. Surprisingly, we see that as the mass grows the error decreases for the pair with larger initial separation.
 As we increase the mass, we expect that the particle is less sensitive to compressibility effects. We also note that for a greater mass, the amplitude of the cycloid for a given initial position is larger. We see the effect of this on the pair with smaller initial separation. The error grows rapidly as the mass approaches $\mathcal{M}=30$ at this point, the particles collide, causing the vortices to annihilate. In general, the model predicts the behaviour of more massive particles more accurately as long as the inequality $\xi \ll R \ll d$ holds. However, increasing the mass causes larger amplitudes, subsequently causing the inequality to break.

In another protocol, we keep the mass fixed at $\mathcal{M}=17.15$ and vary the initial size of the vortex-antivortex pair. The nonlinear shape to this error is again due to competing assumptions of our model. For low separations the finite size of the particle becomes important, thus we see high error. We conjecture that the minimum that follows is due to the change in velocity due to the separation, $(\partial u/\partial d)2R$, becoming small compared to the velocity, see Fig.~\ref{u_du} in Appendix A. The error always stays of the order of a healing length. This clearly demonstrates the usefulness of a simple ODE based MFM model, which is easily solved numerically and able to capture the phenomenological motion of the vortices loaded with particles. Moreover, it predicts their motion to high accuracy.
 

Above examples show that the MFM model provides good description for a simple configuration. Now we extend its use to predict more complex configurations. 

As mentioned earlier, one of the limitations of our model is that it does not capture the possible annihilation of vortices during a collision event. This process plays an essential role in the dynamical evolution of an assembly of vortices in 2D in the presence of particles, which can potentially increase the annihilation rate~\cite{Vish2017}. This has implications for the quench (relaxation) dynamics of the 2D superfluid system, which undergoes the Berezinskii-Kosterlitz-Thouless (BKT)\cite{Berezinskii1971,Kosterlitz_1973} phase transition via a vortex annihilations. Consequently, the increased annihilations in 2D and vortex reconnections in 3D will have a strong influence the velocity statistics in a turbulent state. 

In Fig.~\ref{M21P29} we show that the model describes the dynamics well up until a collision in which the vortices annihilate. The annihilation is followed by linear motion of the particles, with the particles conserving their momentum from the collision. Although the current models do not account for the annihilations, this feature can be added in a ad hoc manner, for example as in Ref.~\cite{Baggaley11}. To add this we require further study of the collisions to define some particle separation cut off value for which the vortices annihilate. In order to define such a cut off we need to study the different possible dynamics during such a collision and to further understand the exchange of momentum of the particles and the role of the acoustic component. However, this is beyond the scope of the current work.



\subsection{More complex configurations}

We studied the case of particles on vortices in a dipole configuration in detail. We now want to briefly consider further configurations to see how the model performs in more complex set-ups.

It is well known that vortices with more than one quanta of circulation are unstable~\cite{Shin2004,Mottonen2003} and move apart due to the emission of sound due to the rotational acceleration. There have been studies on the role of static potentials in stabilising the vortices~\cite{Teles2015StabilityCO,Gustafson:2009aa}; however, this can be extended to cases with external flows and dynamic potentials like the particles presented here. We show that the particles can be used to add stability and to create a localised region in a flow with circulation larger than one quanta.

\begin{figure*}[!htpb]
\center{
 \includegraphics[width=0.35\linewidth]{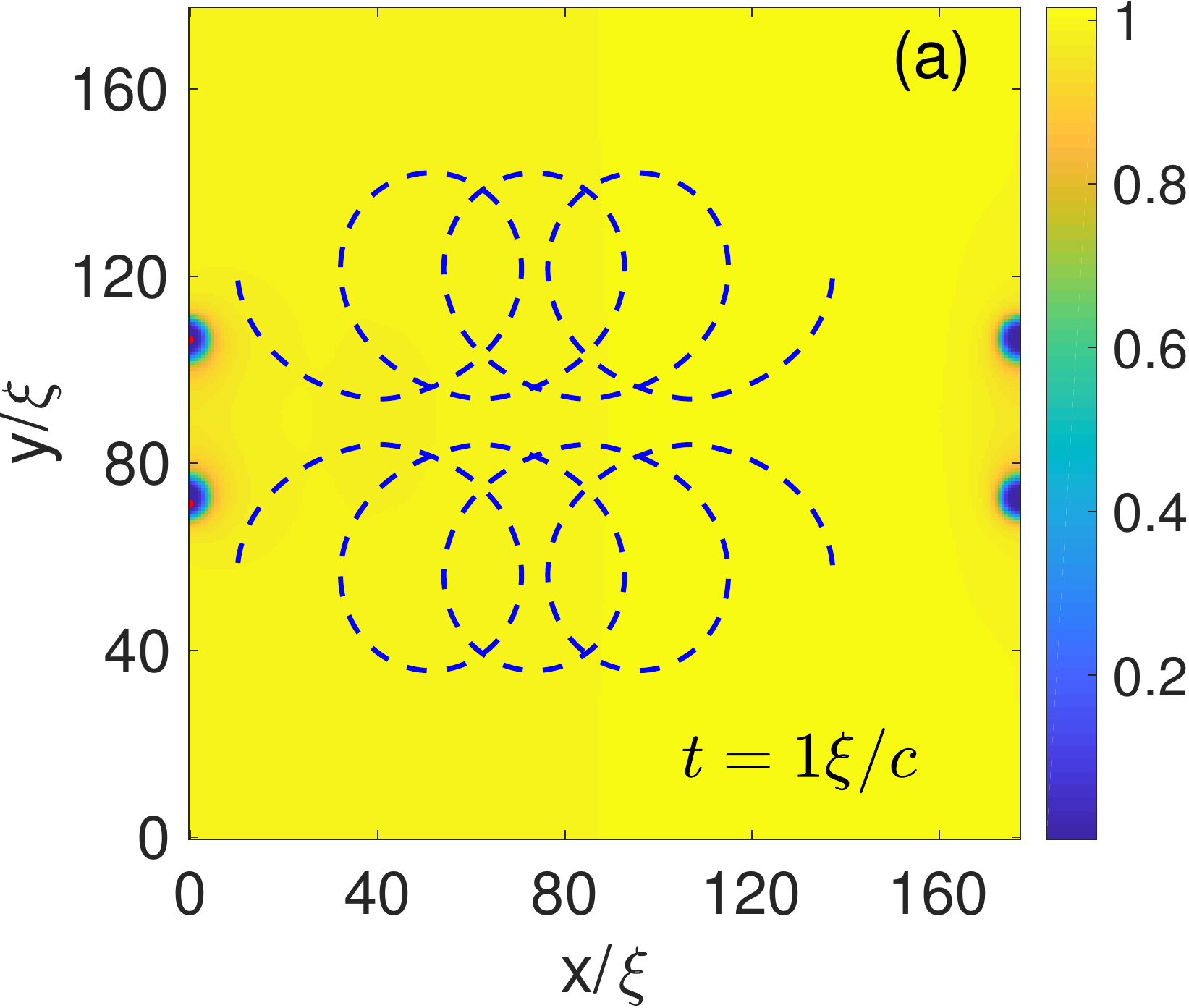} \hspace{6mm}
 \includegraphics[width=0.35\linewidth]{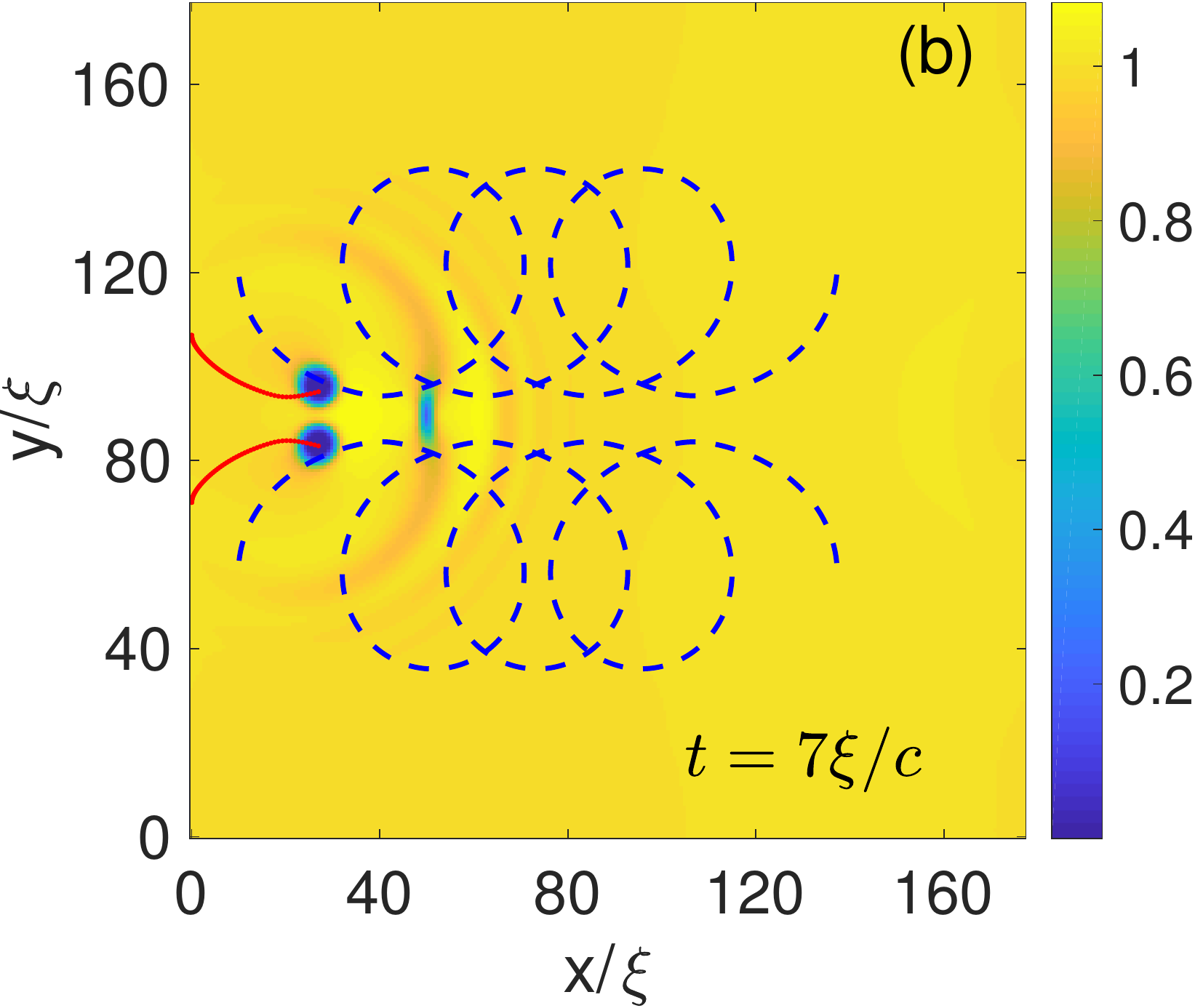}\\
  \includegraphics[width=0.35\linewidth]{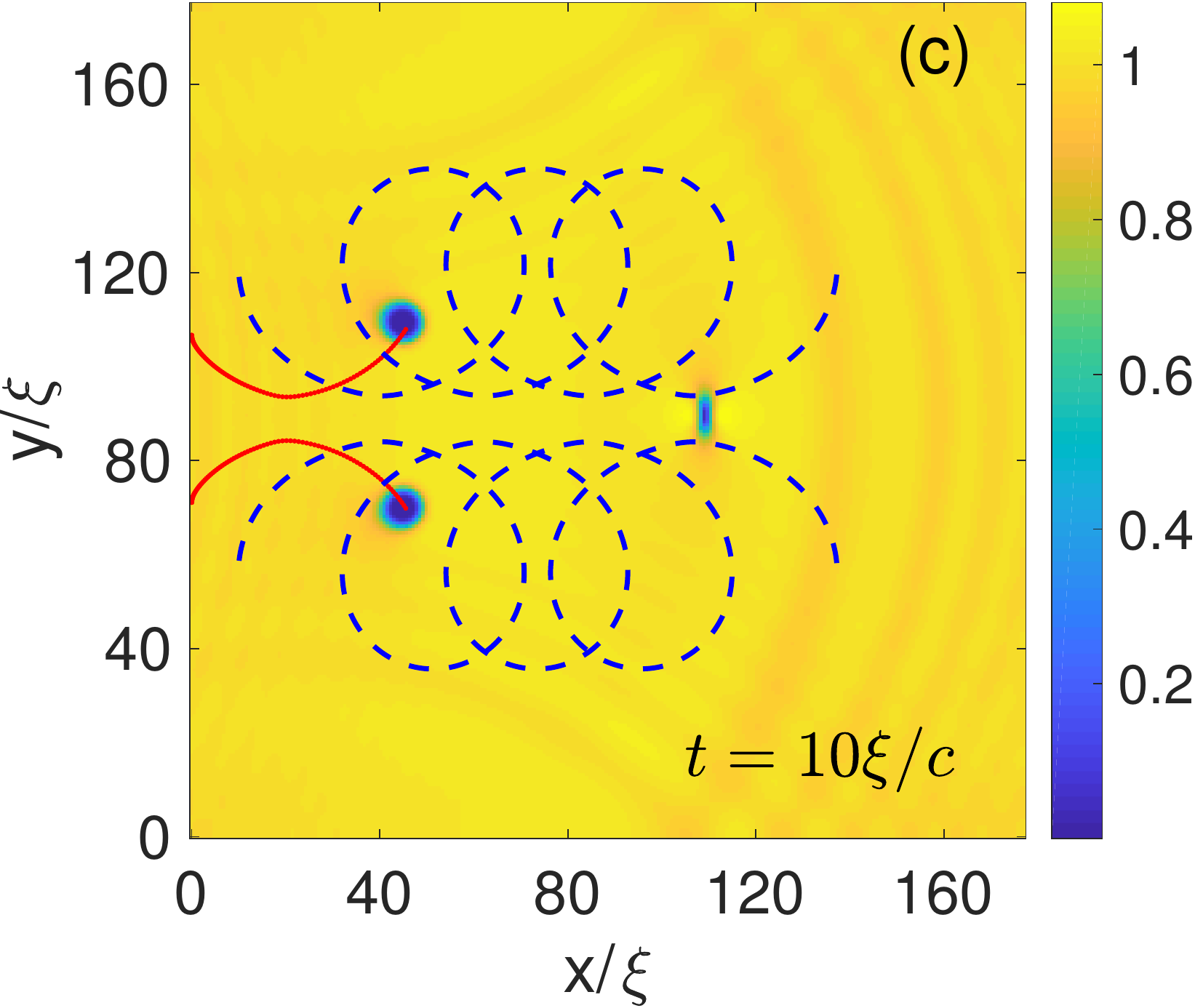}\hspace{6mm}
 \includegraphics[width=0.35\linewidth]{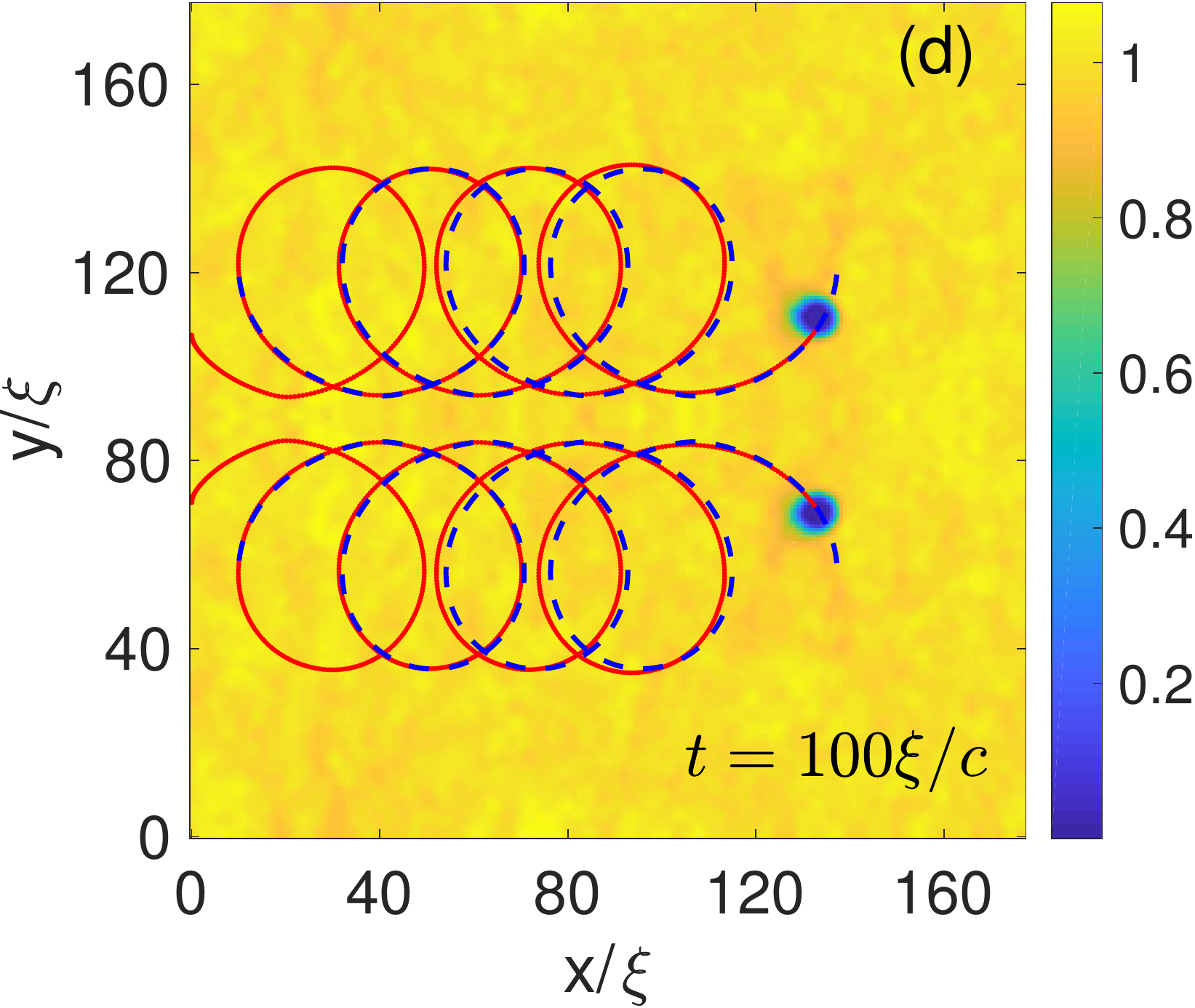}
\caption{{\it (Color online)} Multiple vortices on particles: Psuedo-color density ($|\psi(\mathbf{r},t_f)|^2$) plots with DNS (red) and MFM (blue dashed) trajectories overlaid at different times as presented in the panel. The MFM trajectories are only calculated post-collision. Initially each particle has two vortices or two quanta of circulation on its core. The Robert-Jones soliton is shown by large density change in (b) and (c).}
\label{loop}}
\end{figure*}

 
 We consider a vortex-antivortex pair configuration similar to that in Fig.~\ref{Schem}, but where each vortex has two quanta of circulation and a particle trapped on it. Such a system exhibits an exciting dynamics, it remains stable, i.e. both vortices remain on the particle, up until the collision. We show the collision of the particles in Fig.~\ref{M21P29}, where on colliding only one of the pairs is annihilated via the emission of an unstable Robert-Jones\cite{0305-4470-15-8-036} soliton. The other pair remains on the particles, continuing to propagate, with the dynamics being captured by the MFM. Figure~ \ref{loop} shows that the MFM can successfully model the post-collision dynamics. We initialise the MFM with a position and velocity taken from the simulated GP data after the collision. Even in the presence of considerable background sound waves, we obtain an excellent agreement with the GP dynamics, with the error being $\epsilon=2.74\,\xi$. The fact that the collisional interaction exhibits annihilation of only a single vortex-antivortex pair merits further investigation, as it is not trivial to understand why the entire vortical charge on the particle was not annihilated. This example also emphasises how particles can change the dynamics of a simple configuration, and that reconnection events may be more common when vortices are loaded with particles.

\begin{figure*}[!htpb]
\center{
 \includegraphics[width=0.35\linewidth]{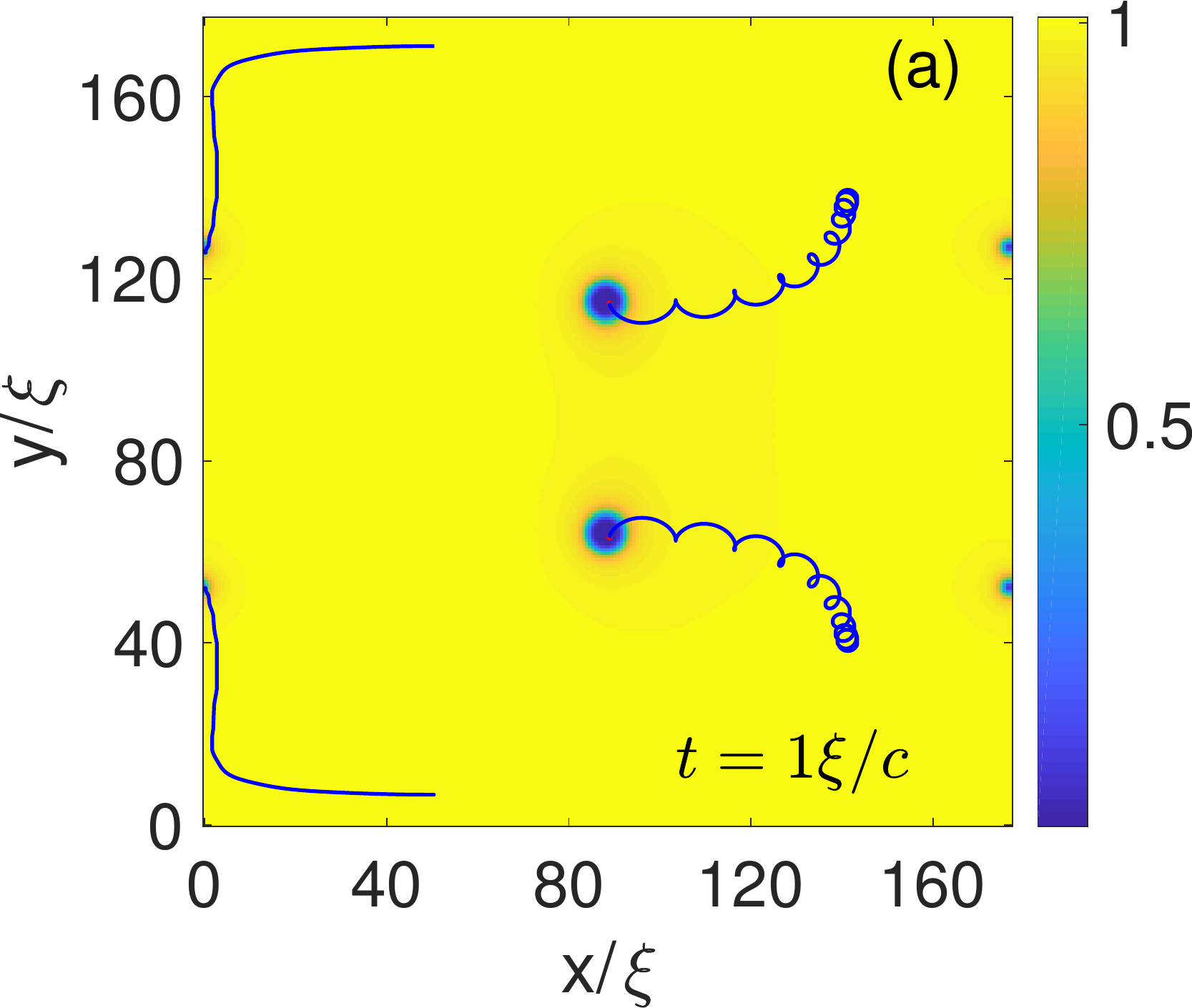}\hspace{6mm}
  \includegraphics[width=0.35\linewidth]{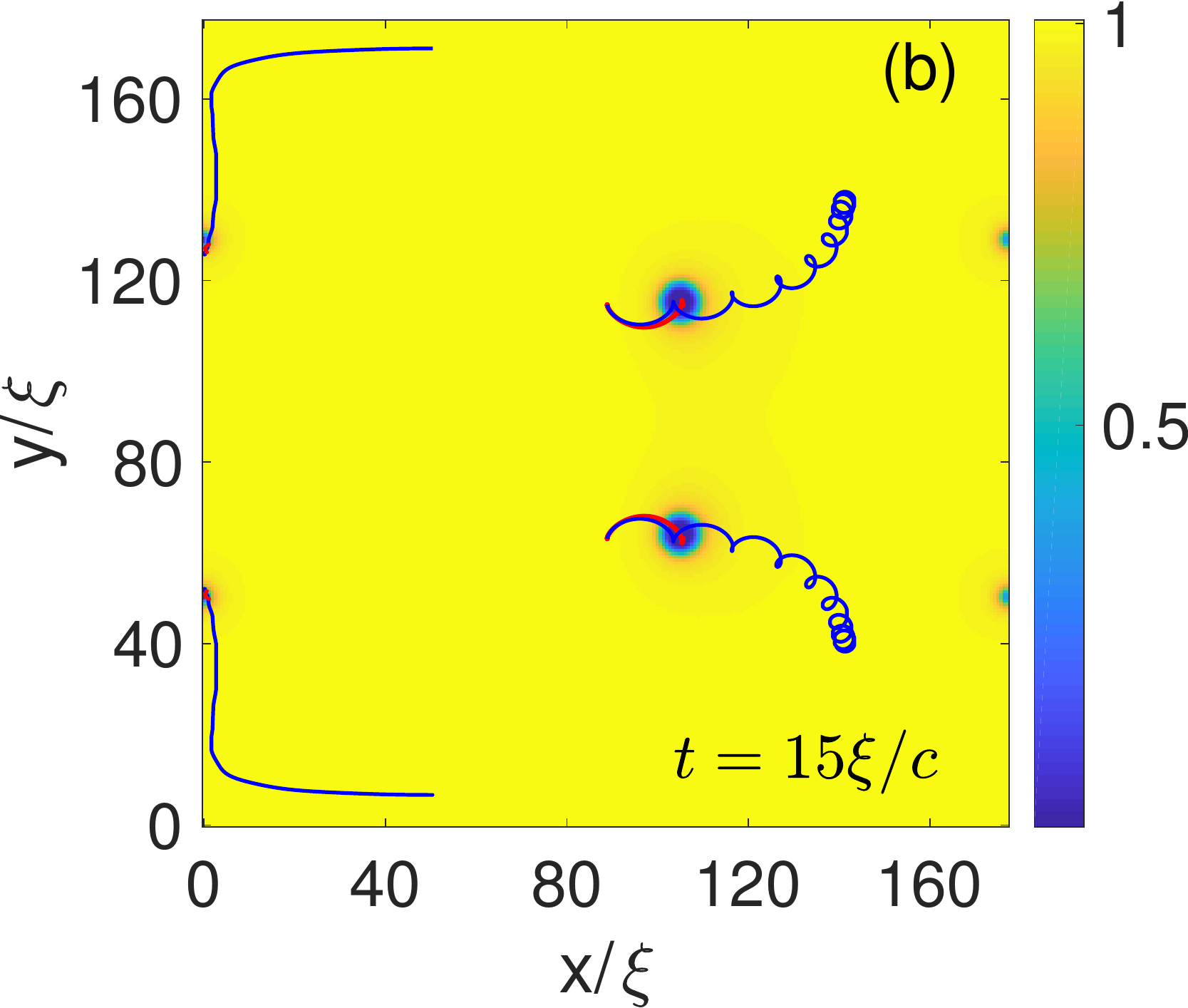}\\
  \includegraphics[width=0.35\linewidth]{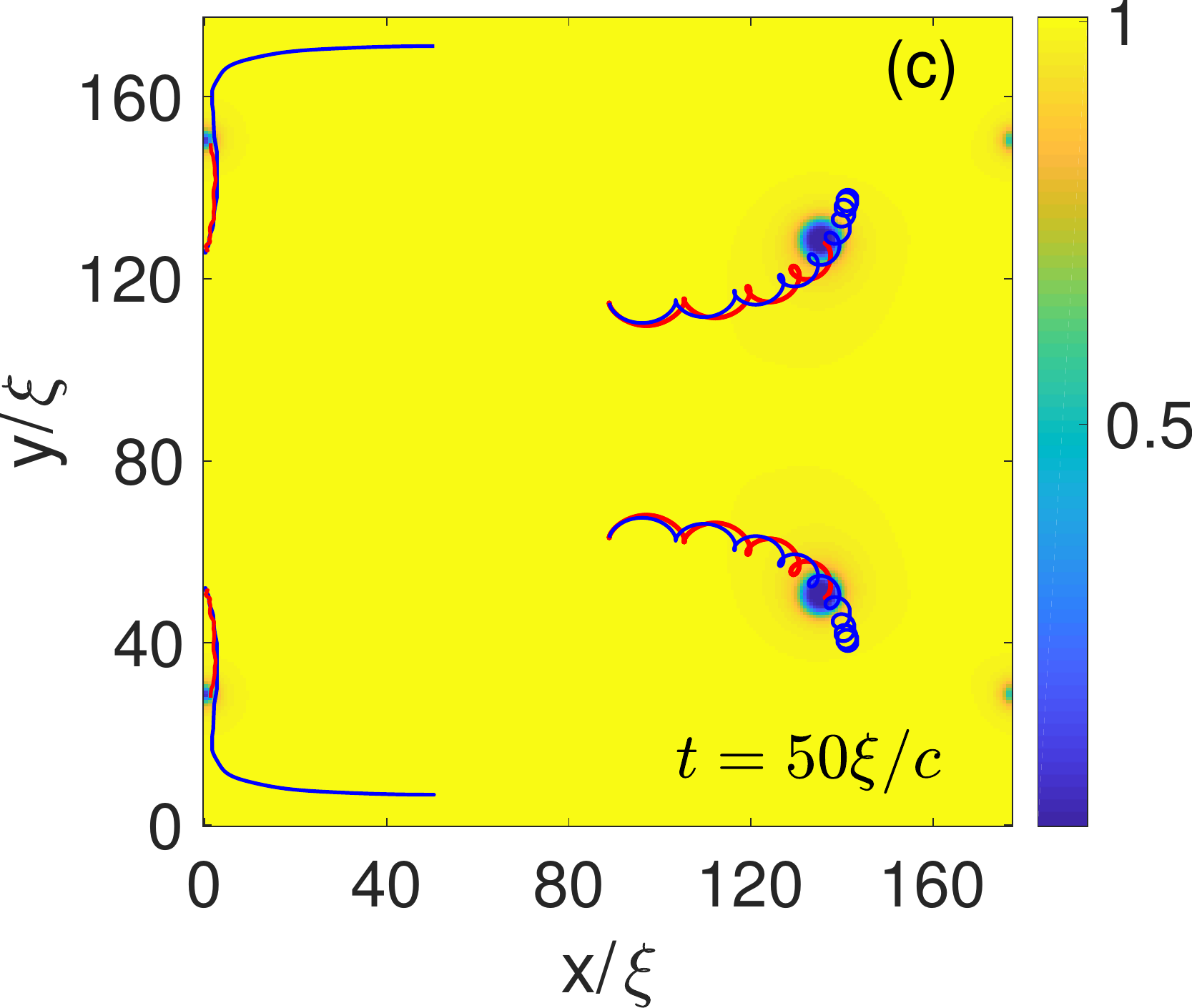}\hspace{6mm}
  \includegraphics[width=0.35\linewidth]{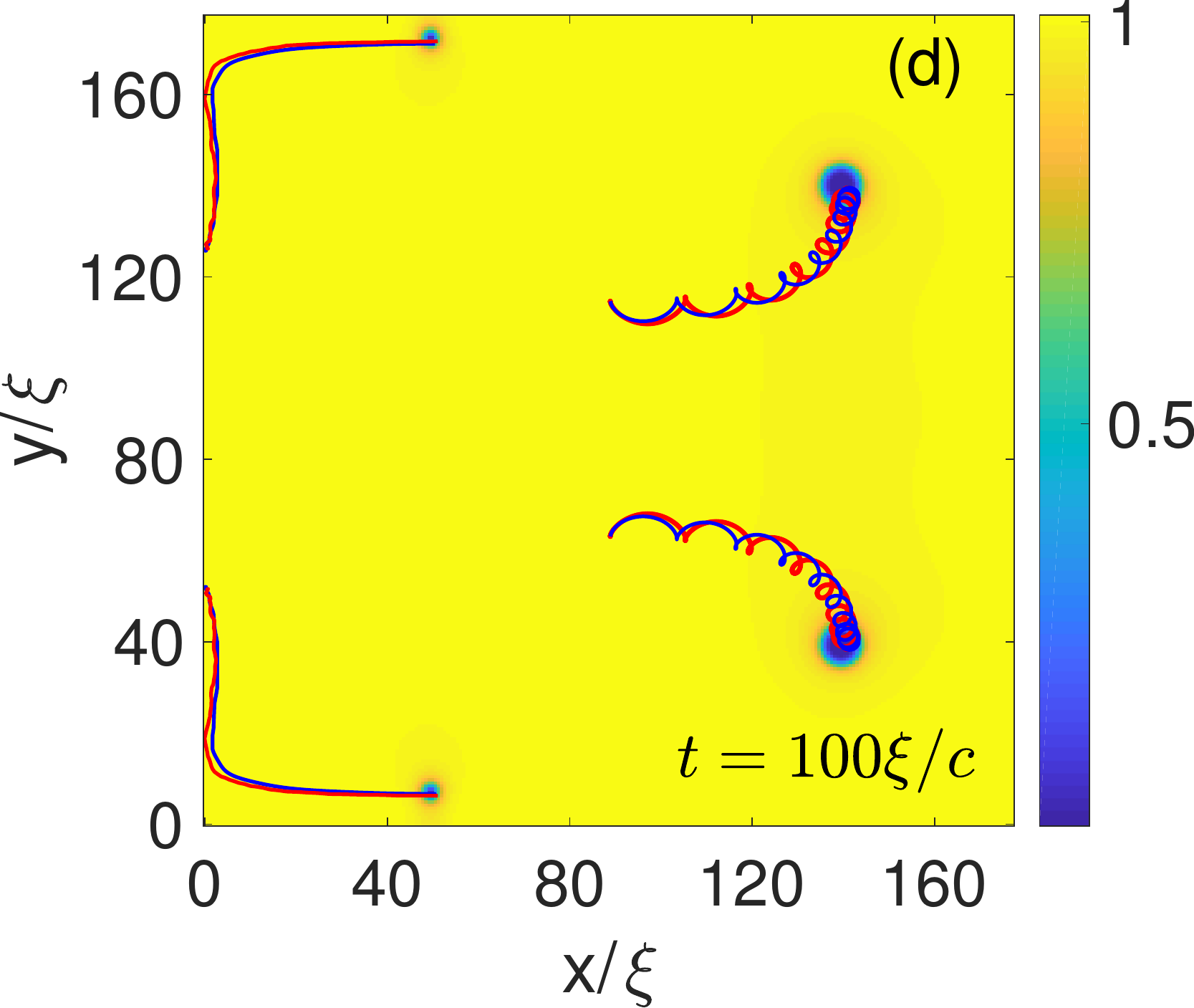}
\caption{{\it (Color online)} Free vortices and loaded vortices. Parameters: mass ratio is $\mathcal{M}=17.15$ and initial separation $d_0=52.5\xi$. Pseudo-color density ($|\psi(\mathbf{r},t_f)|^2$) plots with DNS (red) and PV+MFM (blue) trajectories of particles and free vortices overlaid. The smaller disks in the pseudo-color density plots are free vortices. The DNS (red) and PV+MFM (blue) are overlaid and are difficult to distinguish (see color online).}\label{2v2p1}}
\end{figure*}

To further increase the complexity, we consider a final example in which free vortices (i.e. vortices not loaded with particles) are positioned in front of a vortex-antivortex pair loaded with particles, in the latter each vortex has two quanta of circulation. We compute the motion of the free vortices using \eqref{AMCW} such that the velocity contributions from the vortices loaded with particles are the same as if they were free vortices. The motion of the particles is once again by numerically solving the Eqs.~\eqref{2Dx} and \eqref{2Dy}; however, the velocity field is more complicated now. This model is more general as it can handle a combination of free vortices and vortices trapped on particles; we will refer to this as PV+MFM (Point vortex and Magnus force model). We employ a simple Euler method to advance in time, with the velocity field coupling the two motions. The model predicts well the general motion not only of the particle, and also the free vortices as seen in Fig.~\ref{2v2p1}. The average error of the model is $3.57\xi$, and one can see in FIG.~\ref{2v2p1} that with the additional complexity the dynamics are captured phenomenologically.


\section{Conclusions}

We have shown that the particles trapped on vortex cores experience a Magnus force in the presence of neighbouring vortices or a relative background flow. For simple configurations, where the flow velocity is known, this method can be employed to efficiently predict particle trajectories. It is also possible to generalise our model in the presence of many free vortices by accounting for their motion using the usual point vortex expressions. The MFM predictions are good enough, as long as we can ignore the finite size of the particles, e.g., for large size vortex-antivortex pairs, where it fits well even for massive particles. The model performs well outside of its formal limits of applicability. 

The MFM description naturally extends to 3D by combining the model with the LIA, we hope to further develop the method beyond LIA and incorporate the full BS description in future work. Our model can be generalised to include cases in which annihilation takes place and of course, describes well the post-collision dynamics. Further work could focus on how the attractive forces added to vortices can increase vortex annihilations and further aid phase transitions like the BKT phase transition. The development of a simple theoretical framework which accounts for the motion of particles in the presence of vortices would significantly improve the ability to study the effect of particles in superfluid turbulence.
 We could algorithmically add annihilations of vortices and vortices on particles much using similar methods which are commonly used\cite{Baggaley:2012aa}; this will be the subject of future work. To do this, we would need to choose some physical cut-off value of the particle separation for which vortices annihilate. This could be found by further studies into the annihilation of vortices in the presence of external potentials. However, with the current model we have a fast and accurate way to simulate 2D turbulence interacting with tracer particles.


\begin{acknowledgments}
AG is supported by EPSRC studentship Ref~1642416. SN is supported by the Chaire D’Excellence IDEX (Initiative of Excellence) awarded by Universit\'e de la C\^ote d’Azur, France.
MEB was supported by Agence Nationale de la Recherche through the project QUTE-HPC ANR-18-CE46-0013.
This project has received funding from the European Union’s Horizon 2020 research and innovation programme under the grant agreements No 823937 in the framework of Marie Skłodowska-Curie HALT project and No 820392 in the FET Flagships PhoQuS project.
 Computations were carried out on the High Performance Computing Cluster and supported by the Scientific Computing Research Technology Platform at the University of Warwick. 
\end{acknowledgments}

\appendix

\section{Weiss-McWilliams formulae for ideal fluids and GP system} \label{DV}

The adaptation of the Weiss-McWilliams formula for the translational velocity of a vortex-antivortex pair in an ideal fluid to the GPE system with a periodic domain requires a modification because of the periodicity of the phase of the wave function in the latter. Below we demonstrate this for a vortex-antivortex pair, which we then generalize to the case of $N$-pairs.

\begin{figure}[!h]
\includegraphics[width=1.0\linewidth]{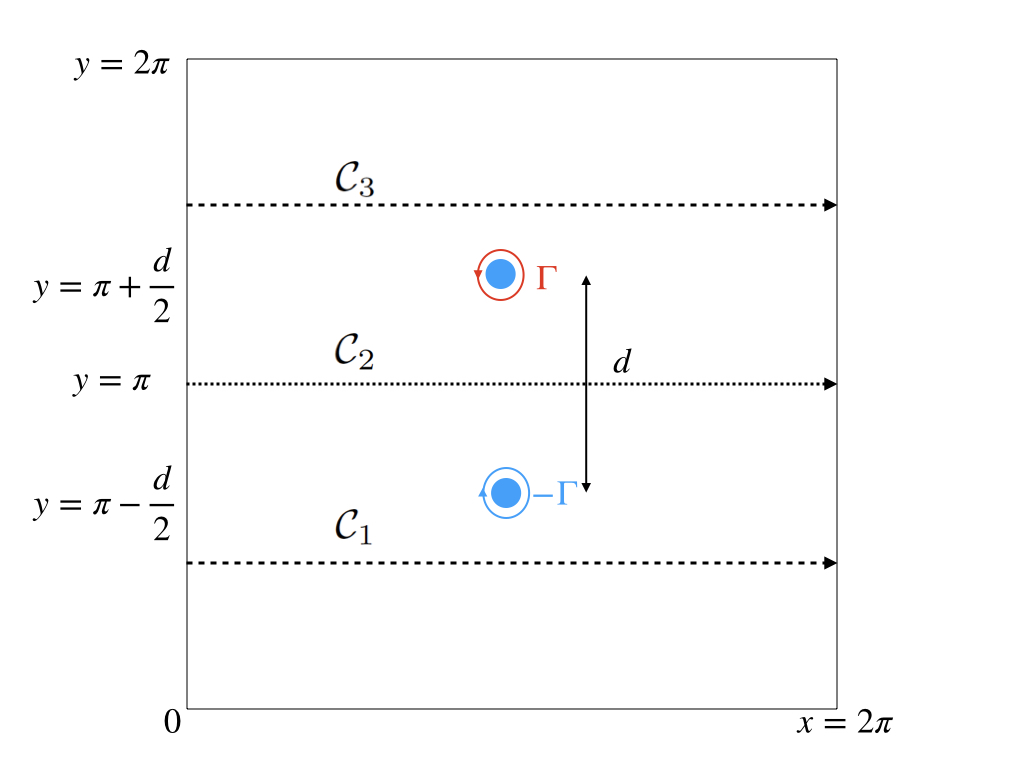}
\caption{{\it (Color online)} Schematic of two vortices separated by distance $d$. Overlaid with the contours on which we calculate the $x$-contributions of the circulation.
}\label{fig:SCHEM2}
\end{figure}

Consider a vortex-antivortex pair, translating along the $x$-axis with vortex at $(x,\pi+d/2)$ and antivortex at $(x,\pi-d/2)$. Now let us consider the circulation along three lines $\mathcal{C}_i$ for $i=1,2,3$, as shown in Fig.~\ref{fig:SCHEM2}:
\begin{align}
\mathcal{C}_i(d) = \int_{0}^{2\pi} u(x,y_i)dx,
\end{align}

where $0 < d < 2\pi $; $0 \leq y_1 < \pi-d/2$; $\pi-d/2 < y_2 < \pi+d/2$ and $\pi+d/2 < y_3 \leq 2\pi$. These contours can be closed by joining $x=0$ and $x=2\pi$, without generating any contribution to the circulation, as guaranteed by the periodic boundary condition. Thence,
\begin{align}
\mathcal{C}_1 - \mathcal{C}_2 = -\Gamma, \\
\mathcal{C}_2 - \mathcal{C}_3 = \Gamma, \\	
\mathcal{C}_1 = \mathcal{C}_2-\Gamma = \mathcal{C}_3,\label{c1isc3}
\end{align}
for the contours enclosing vortices. The circulation can now be solely expressed as a function of the $y$-position of the contour and the separation between the vortices:
\begin{align}
\int_0^{2\pi} u(x,y) dx = \mathcal{C}(y,d)= \int_0^{2\pi} - \frac{\partial h}{\partial y} dx
\end{align}
where $h$ is the stream function and $u=-\partial h/\partial y$. Let us define the mean velocity
\begin{align}
\bar{u} & =\frac{1}{(2\pi)^2}\int_0^{2\pi} \mathcal{C}(y,d) dy \\
& = \frac{1}{(2\pi)^2}\int_0^{2\pi}\int_0^{2\pi} u(x,y) dx dy .\label{contint}
\end{align}
From the periodicity of stream function it follows that the mean velocity is zero.
The integral in \eqref{contint} can be then computed as a sum of the areas multiplied by the circulation, that is,
\begin{align}
 \left(\pi-\frac{d}{2}\right)\mathcal{C}_1 +\mathcal{C}_2 d +\left(\pi-\frac{d}{2}\right)\mathcal{C}_3 & = 2\pi\mathcal{C}_1+\Gamma d \nonumber\\
 & =0,
\end{align}
by using equation \eqref{c1isc3}. Thus $\mathcal{C}_1 = -\Gamma d/(2\pi)$ for a system with a periodic stream function. Note that the Euler equations are Galilean invariant and any constant velocity can be added to the system which corresponds to moving to a different inertial frame. However, the requirement that the streamfunction is periodic fixes the frame of reference such that the mean velocity is zero. This is the choice of frame in Ref.~\cite{Weiss:1991aa}. Such a choice is inconsistent with the periodicity of the wave function, as it would result in a phase change which is not an integer multiple of $2\pi$ (for $\Gamma=2\pi, \mathcal{C}_1=d \neq 2\pi n$). Therefore, if we want to use the Weiss-McWilliams prescribed velocity (ideal fluid case), we must work in a frame in which the phase of the wave function is periodic. Thus, we must add a constant background velocity equal to $\mathcal{C}_1/2\pi$.

Note that in order to be consistent with the contents of the main text, in our discussion here, we have used a vortex-antivortex pair that is only separated along the $y$-direction. However, our arguments are still valid if the vortices are separated in the $x$-direction as well. In which case, we simply have to repeat the argument with vertical contours which separate the vortices to find the contribution to the $y$-component of the velocity.

Moreover, our discussion here is generalisable to the case of $2N$ vortices, with an equal number of vortices and antivortices. This is so because the system is linear, i.e., the contribution to the velocity field is additive. Also, the argument is valid, as the sum of the periodic phases will also be periodic. Therefore, for a system of $2N$ vortices, the additional background velocity $\mathbf{u_b}$ is
\begin{align}
\mathbf{u_b}=\sum_i^{2N} \frac{\Gamma_i}{(2\pi)^2}\mathbf{\hat{e}}_z\times\mathbf{x}_i,
\end{align}
where $\mathbf{x}_i$ is the position of the $ith$ vortex, which is, up to a constant, the total momentum of the point vortex system.


For an odd number of vortices, the system will have a net circulation. Therefore, to make arguments as, we now have to work in a rotating frame of reference. This is same as adding a constant background vorticity, such that the frame has net circulation zero. The ratio between the angular velocity of the rotating frame and the vorticity is a half. Other valid configurations could include constant vorticity such as shear. In this case, such formulation is consistent when periodicity is imposed in sheared coordinates.

\begin{figure}[!t]
\includegraphics[width=1.0\linewidth]{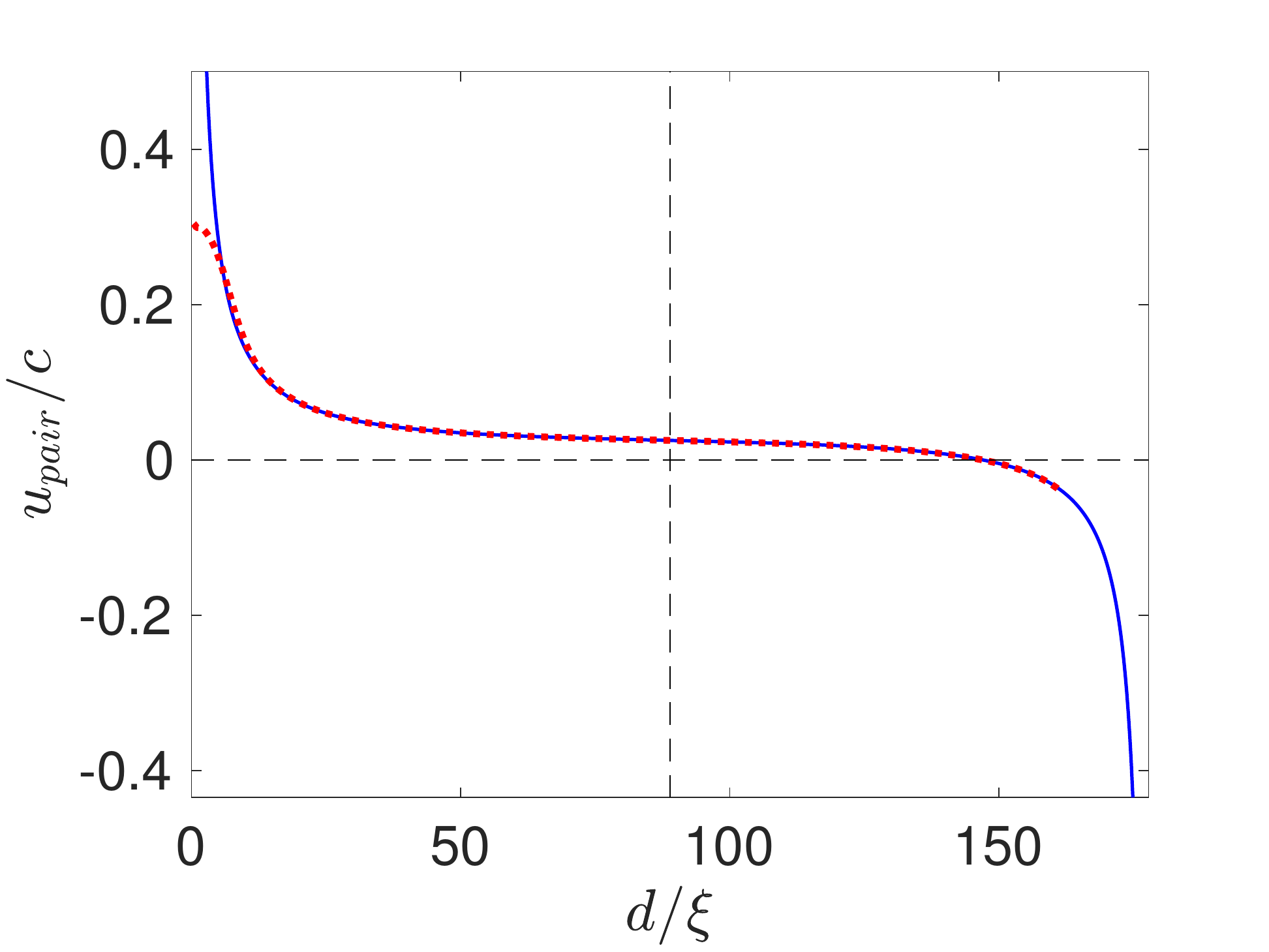}
\caption{{\it (Color online)} Velocity of the vortex-antivortex pair as a function of its size. The red dashed curve represent the data obtained from the GPE simulations, whereas the blue solid line indicates the velocity given by the adapted Weiss-McWilliams formula, see text for more details. The black dashed horizontal line indicates $u_{\rm pair}=0$ and the black dashed vertical line marks the mid point of the periodic domain.
}\label{fig:0}
\end{figure}

In Fig.~\ref{fig:0} we show the comparison of the vortex-antivortex pair velocity as obtained from the GPE DNS (dashed red curve) and the adapted Weiss-McWilliams formula (blue curve). For very small pair size, we observe a disagreement, as now the vortex core size is important. At still very small sizes, vortices annihilate and become a localised density perturbations. The original Weiss-Mcwilliams (ideal fluid) x-directional velocity of a vortex pair (as in \ref{fig:SCHEM2}) is given by 

\begin{align}
u_{WMc}(d)=-\frac{\Gamma}{4\pi}\sum_{n=-\infty}^{\infty}\frac{\sin(d)}{\cosh(2\pi n) - cos(d)} .\label{AMCW}
\end{align}
Therefore, the expression obtained by adapting this to the case of the GPE in a periodic domain is given by
\begin{align}
u_{pair}(d) = -\frac{\Gamma}{4\pi}\left(\sum_{n=-\infty}^{\infty} \frac{\sin(d)}{\cosh(2\pi n) - cos(d)} +\frac{d}{\pi}\right).\label{AMCW1}
\end{align}

For our purposes, we calculate the velocity until it has converged within $10^{-9}$, this corresponds to retaining $9$ terms, i.e. $n=-4$ to $n=4$. Therefore, to the leading order
\begin{align}
u_{pair}(d)=\frac{\Gamma}{4\pi}\left(\frac{1+\cos(d)}{\sin(d)} +\frac{d}{\pi}\right),
\end{align}
and its derivative with respect to separation is 
\begin{align}
\frac{\partial u_{pair}(d)}{\partial d}=-\frac{\Gamma}{4\pi}\left(\frac{1+\cos(d)}{\sin^2(d)} +\frac{1}{\pi}\right).
\end{align}

This has a minimum at $d=\pi$, thus the approximation of constant velocity across the particle is most suitable close to half the domain size, $L/2$. 

\begin{figure}[h]
\center{
 \includegraphics[width=0.8\linewidth]{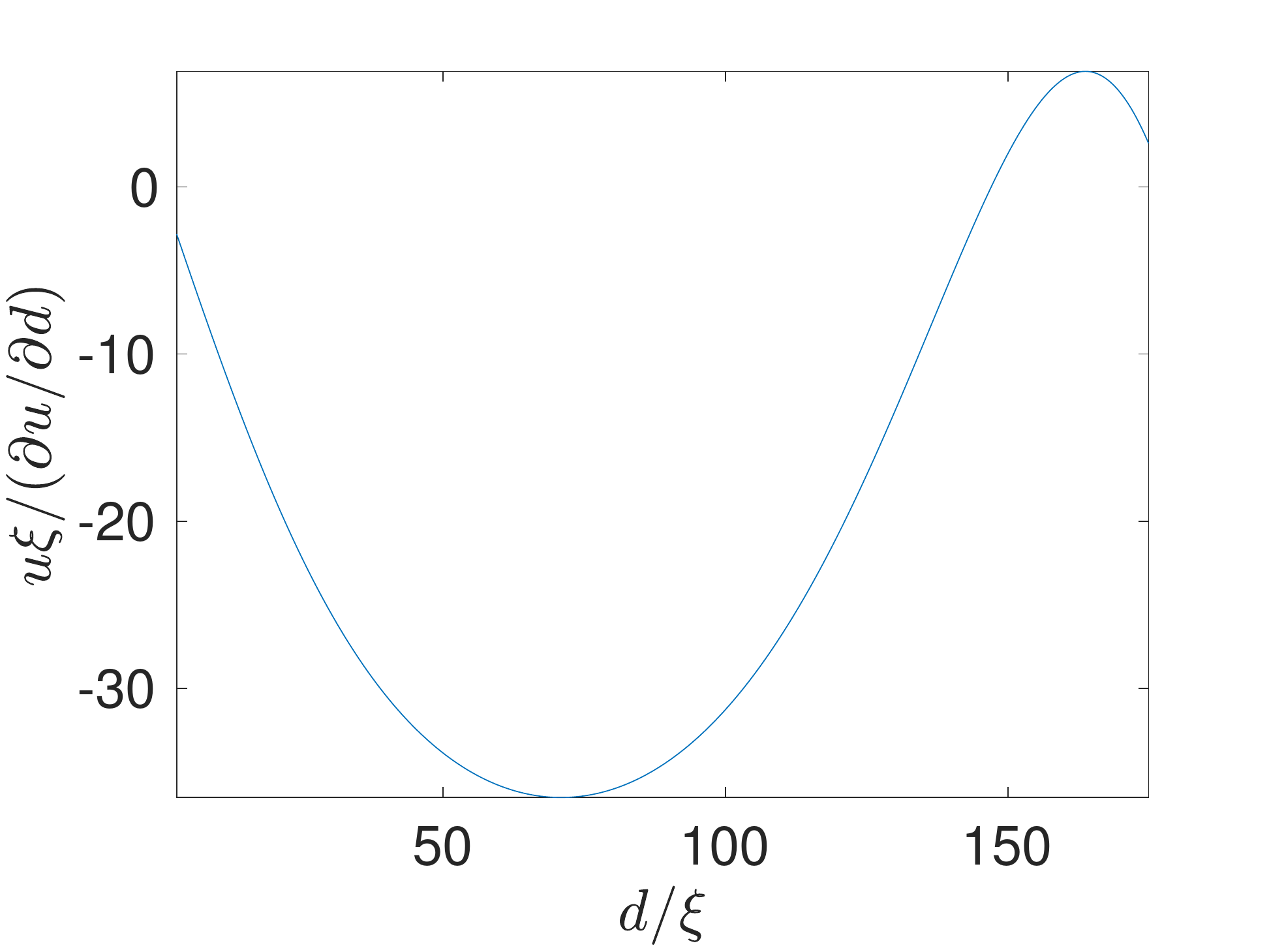}
\caption{Derivative of velocity with respect to separation against separation.}\label{u_du}}
\end{figure}

\bibliographystyle{phjcp}
\bibliography{ref}

\end{document}